\documentclass[a4paper,11pt]{article}
\pdfoutput=1 % if you are submitting a pdflatex (i.e. if you have
\usepackage{jinstpub} % for  details on the use of the package, please
\usepackage{hyperref}
\usepackage{lineno,hyperref}
\usepackage{amsmath}
\usepackage[utf8]{inputenc} % To be able to write ⌀ in the first place
\usepackage{graphicx}
\usepackage{subcaption}

\title{\boldmath Testing a large size triple GEM detector for the first station of the CBM-Muon Chambers with a high-intensity gamma source at GIF++ under large-area illumination}

\author[a,b,1]{Apar Agarwal, \note{Corresponding author.}}
\author[a,b]{Souvik Chattopadhay,}
\author[a,b]{Pawan Kumar Sharma,}
\author[a]{Anand Kumar Dubey,}
\author[a]{Jogender Saini,}
\author[a]{Vikas Singhal,}
\author[a]{Vinod Negi,}
\author[a]{Ekata Nandy,}
\author[a]{Chandrasekhar Ghosh,}
\author[c]{David Emschermann,}
\author[a,b]{Zubayer Ahammed,}
\author[a]{Subhasis Chattopadhyay}

\affiliation[a]{Variable Energy Cyclotron Centre, Kolkata, INDIA}
\affiliation[b]{Homi Bhabha National Institute, Mumbai, INDIA}
\affiliation[c]{GSI Helmholtzzentrum für Schwerionenforschung, Darmstadt , Germany}

\emailAdd{a.agarwal@vecc.gov.in}
\abstract{The physics studies at heavy-ion nucleus-nucleus collision experiments demand reliable detectors at high particle flux. Therefore, Gas Electron Multipliers (GEM) detectors, which show resilience to extreme radiation, are one of the prime choices for the upcoming Compressed Baryonic Matter (CBM) experiment at the Facility of Antiproton and Ion Research, Germany. However, operating them under these demanding conditions requires a systemic study at the highest incident particle flux. To this end, we have conducted extensive tests on a real-size triple GEM detector module with the high-intensity gamma flux using the Cs-137 source at the upgraded Gamma Irradiation Facility (GIF++) at Conseil Européen pour la Recherche Nucléaire (CERN). The detector response, particularly regarding the gain and efficiency of muon detection, was studied extensively with and without a gamma source in a free-streaming mode using self-triggered electronics. This configuration will be necessary for the CBM experiment since it will observe unprecedented event rates of about '10 MHz' for Au$-$Au collisions. The analysis reveals an alignment between the expected and observed value of gain and efficiency with an increasing intensity of gamma flux at the operating voltage. The test results demonstrate that the large-size GEM detector prototype can handle elevated gamma rates of approximately '17.25 MHz/cm$^{2}$' without significantly impacting its performance or suffering irreversible damage.
}


\begin{document}
\flushbottom
\maketitle

%Compressed Baryonic Matter ( (Facility for Antiproton and Ion Research) 

%\linenumbers

\section{Introduction}
\label{sec:introduction}

Gas Electron Multipliers (GEM) \cite{Sauli:2016eeu} are a type of Micro-Pattern Gas Detector (MPGD) known for their ability to meet large-area requirements, offer design versatility, require minimal maintenance, resist aging, and deliver exceptional performance in high-radiation environments. Due to their specific characteristics when compared to other conventional gas detectors such as the Multi-Wire Proportional Chambers (MWPCs) \cite{Charpak:1968kd}, GEMs are not only being used in various running experiments \cite{Lippmann:2014lay} \cite{Colaleo:2015vsq} but also chosen to be employed in many future particle physics experiments \cite{BARBOSA2019162356} \cite{TSIONOU2017309}. One such future experiment is Compressed Baryonic Matter (CBM) \cite{Cbm} at the upcoming Facility for Antiproton and Ion Research (FAIR) \cite{Fair} in Darmstadt, Germany. Once completed, CBM will boast heavy-ion collisions at a rate of '10 MHz', the highest ever. Such high interaction rates will allow us to study several rare probes with unprecedented precision in the high baryonic density ($\mu_{B}$) region of the Quantum Chromo-Dynamics (QCD) phase diagram. Under these conditions, we require detectors that can perform reliably in a high particle flux. The experiment will be operated in a trigger-less or free streaming mode \cite{Korolko:2017fbz}\cite{DUBEY201462}.
Large-sized triple GEM detectors will be employed in the first two (out of four) stations of the Muon Chambers (MuCh) of CBM \cite{DUBEY2013418} primarily for the measurement of dimuon signals resulting from the decay of the Low Mass Vector Mesons (LMVMs), J/$\Psi$ among others, from the highly dense matter produced during heavy-ion collisions \cite{MuChTDR:161297}. 

As part of the detector development for the CBM, we have tested a prototype module of a large-size triple GEM detector \cite{ADAK201729} together with a bakelite RPC module in the upgraded Gamma Irradiation Facility (GIF++) \cite{Agosteo:2000mk}\cite{H4Intensity} in the Conseil Européen pour la Recherche Nucléaire (CERN) \cite{CERN}. This facility accommodates a Cs-137 source of strength, '11.34 TBq' (activity estimated on the date of the experiment). It can simultaneously illuminate detectors with a muon beam, whose typical intensity is of the order of 10$^{4}$ per spill (or 4.6 seconds). 
The facility provides a suitable environment for measuring several detector characteristics using incident muons amidst high levels of gamma radiation \cite{Pfeiffer:2016hnl}. 
This paper provides a comprehensive analysis of the studies conducted during the September 2023 GIF++ experiment. The primary objective of this work is to evaluate the rate-handling capabilities of our large-area GEM detectors. The findings presented here offer valuable insights that will guide the optimization of operational parameters for GEM detector modules in future experiments, especially those to be conducted in a high gamma radiation environment.

% \begin{figure}[htbp]
% \centering 
% %\includegraphics[width=4.0cm,height=6.0cm]{}
% %\hspace{5mm}
% %\includegraphics[width=4.0cm,height=7.0cm]{Picture_GIF++.png}
% \includegraphics[width=10.0cm,height=6.0cm]{Images/Picture_GIF++.png}
% \caption{\label{fig:mCBMSetupPicture} Side and upstream view of one of the detector setups for muon chambers test at GIF++.}
% \end{figure}

\section{Detector design, readout and the experimental setup}
\label{sec:detector_design}

The triple GEM detector module utilized for the tests at GIF++ was fabricated using a 3-2-2-2 gas gap configuration, denoting the drift gap, two transfer gaps, and an induction gap of '3 mm', '2 mm', '2 mm', and '2 mm', respectively. This module stood out because, aside from the single-masked GEM foils, all of its components were manufactured in India. The module was assembled in a Clean Room Facility (ISO-6) at the Variable Energy Cyclotron Center (VECC), Kolkata. The GEM foils used were of the "single-mask" type, procured from CERN, and were designed according to the specifications discussed in~\cite{Kumar:2021xpj}. 

\begin{figure}[htbp]
\centering 
\subfloat[]{{\includegraphics[width=4.0cm, height=6.0cm]{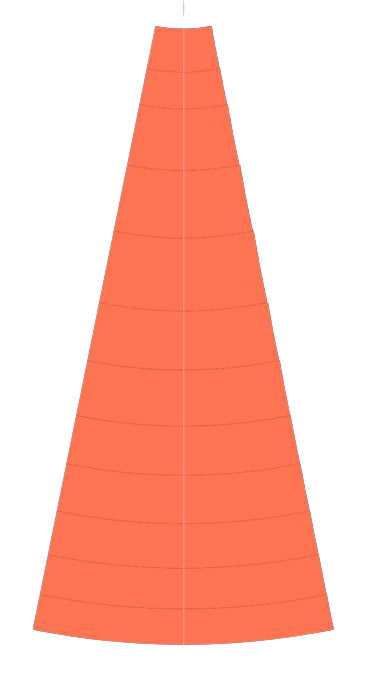}}}
\qquad
\subfloat[]{{\includegraphics[width=5.5cm,height=6.0cm]{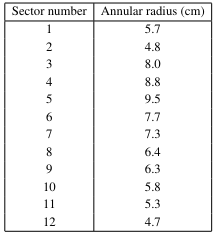}}}
\qquad
\subfloat[]{{\includegraphics[width=3.1cm,height=6.0cm]{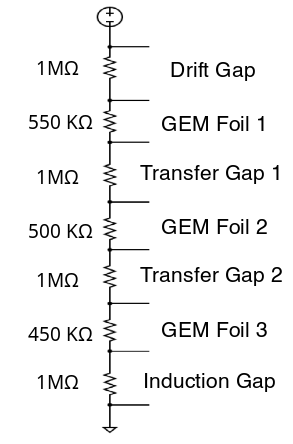}}}
\caption{\textbf{(a)}: Segmentation of GEM Foil into 24 partitions \textbf{(b)}: Annular radial dimension of foil segments \textbf{(c)}: High Voltage divider circuit in GEM module}
\label{fig:resistive_scheme} 
\end{figure}

In general, large-size GEM foils are segmented to minimize capacitive load. At CBM, different regions of the GEM detector module would observe different particle fluxes, with the narrow region (being closer to the beam pipe) facing the maximum particle flux. Therefore, the GEM foil was partitioned as shown in Fig.\ref{fig:resistive_scheme}(a)(b) into 24 segments of varying areas such that each partition collects a comparable amount of charge. The foils were stretched during module assembly using the "pull-out" technique \cite{CMSMuon:2018wlc}, consistent with the procedure employed in earlier versions of the full-size prototypes \cite{Kumar:2021qyi}. 

As part of the Quality Control (QC) process, several tests were conducted on various GEM chamber components, including GEM foils and their mechanical parts such as spacers and supports. Before assembly, all detector components were manually inspected to ensure that their dimensions matched the specified requirements. After assembly, a "gas-tightness" test was performed to verify the absence of leaks through screws and joints.

For the QC of the GEM foils, we individually measured the leakage currents for the 24 segments of each of the three GEM foils of the detector. These measurements were carried out in an ISO-6 category clean room at a temperature of approximately '22$^{\circ}$C', in the air with a relative humidity of approximately 50\%. Only foils with leakage currents below '5 nA' at an applied bias voltage, $\Delta$V = 550 V for every segment, were approved for the final assembly.

\begin{figure}[htbp]
\centering 
\subfloat[]{{\includegraphics[width=6cm, height=3.1cm, angle=-270]{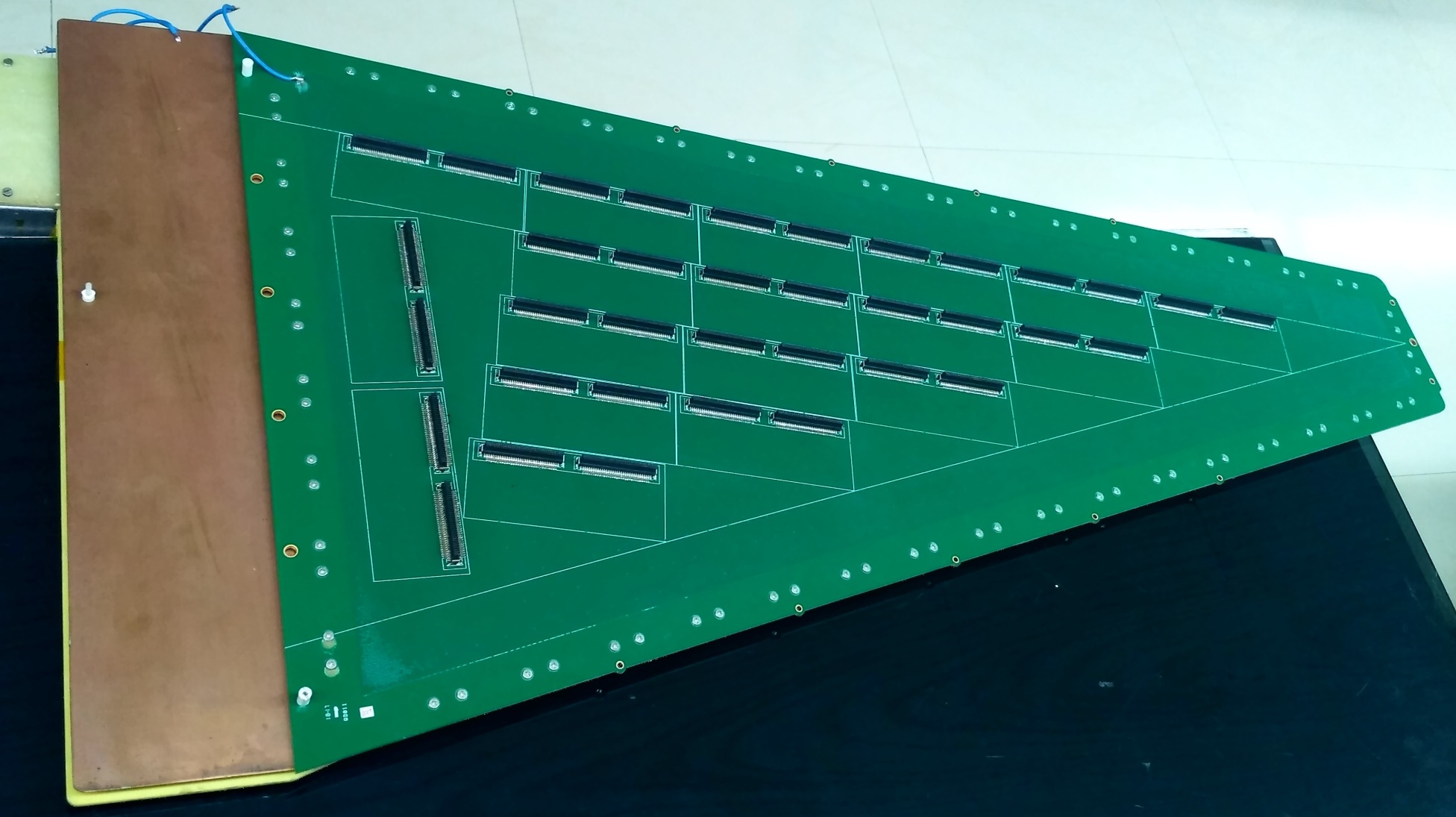}}}
\qquad
\subfloat[]{{\includegraphics[width=3.1cm,height=6.0cm]{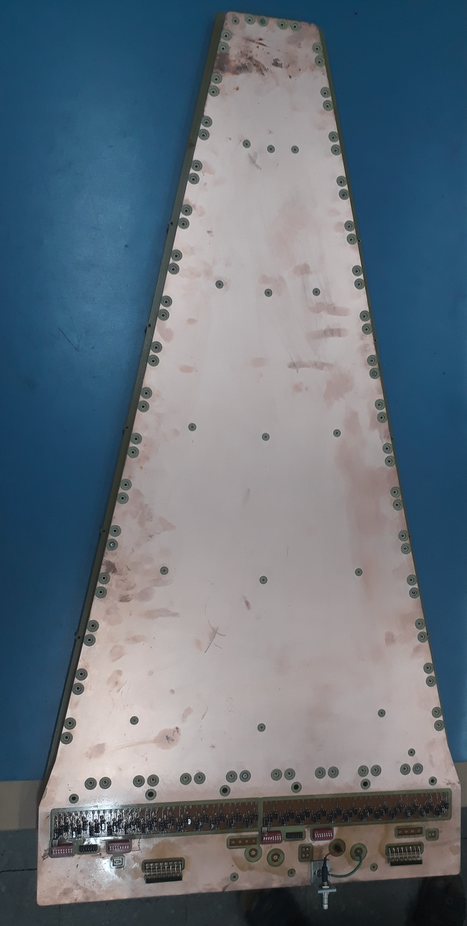}}}
\qquad
\subfloat[]{{\includegraphics[width=5.2cm,height=6.0cm]{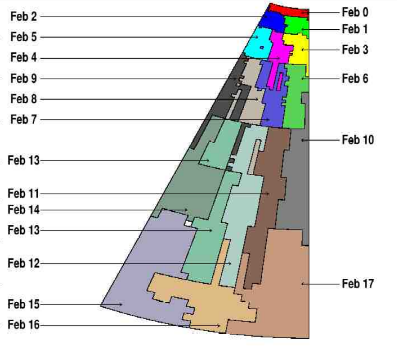}}}
\caption{\textbf{(a)}: GEM module with the readout plane facing upwards. \textbf{(b)}: GEM module with the drift plane facing upwards. \textbf{(c)}: The zones of pads covered by the respective FEB connector number (or FEB number).}
\label{fig:readout_pads_with_dimensions} 
\end{figure}
 
% \section{Data Acquisition System (DAQ)}
% \label{sec:data_acquisition_system_daq}
The active area of the real-size GEM chambers is in the shape of a sector with narrower and broader arcs having chord lengths of '12.1 cm' and '47.0 cm', respectively, while the height is '87.25 cm'. The detector readout plane consists of 2231 trapezoidal-shaped pads (grid of 23 $\times$ 97) with areas progressively increasing from about '0.1~cm$^{2}$' to '2.89~cm$^{2}$'. An array of 18 Front-end Electronic Boards (FEBs), each consisting of 128 electronic channels, \cite{CBM-MuCh:2023gdx} is used to collect signals from the entire active area of the GEM module, as shown in Fig.\ref{fig:readout_pads_with_dimensions} (a and b). Each "pad" of the detector is uniquely connected to an electronic "channel." In a self-triggered system, for any channel (or pad), the signal amplitude from the detector is recorded in digitized form along with the time stamp whenever it exceeds a set threshold. This recorded signal(s) from a pad(s) (above the threshold) is commonly referred to as a digi(s). 

\begin{figure}[htbp]
\centering 
\includegraphics[width=15.0cm,height=6.0cm]{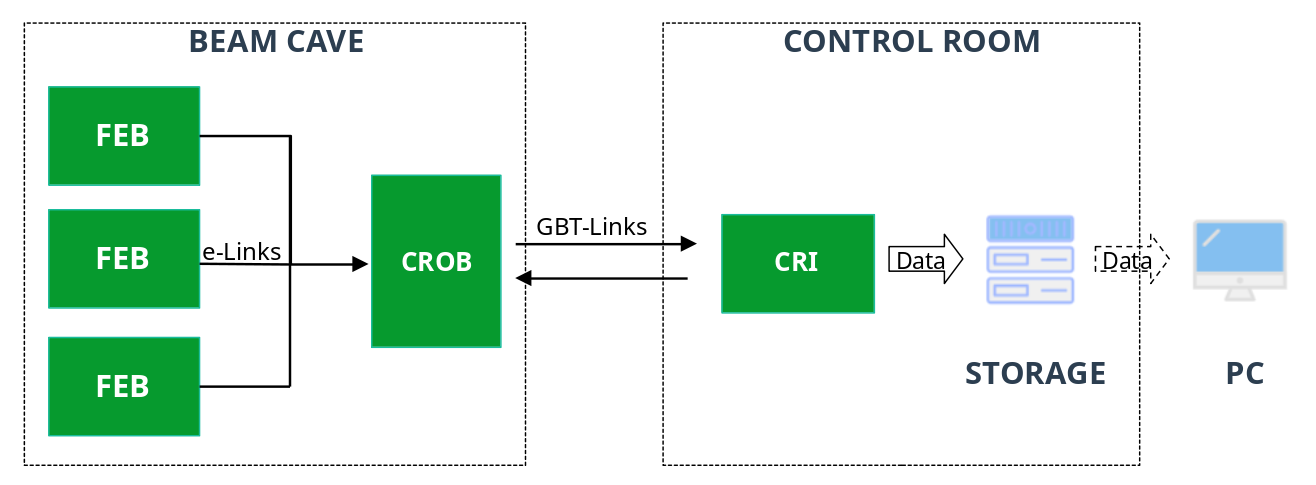}
\caption{\label{fig:DAQ_flowchart} A flowchart depicting the data acquisition system.}
\end{figure}

The area of the active region is measured to be approximately '1900 cm$^{2}$'. In Fig.\ref{fig:readout_pads_with_dimensions} (c), the mapping of the readout pad areas corresponding to the respective FEB numbers has been highlighted. Low-Voltage Distribution Boards (LVDBs) (also developed and fabricated in India) provided a power supply of +12 V to the FEBs and +5 V to the optocouplers through remote operation. Data were collected using STS/MuCh XYTER (version-C) \cite{CBM-SMX} electronics coupled to a mobile DAQ, whose schematic is shown in Fig.\ref{fig:DAQ_flowchart}.

\begin{figure}[htbp]
\centering 
\subfloat[]{{\includegraphics[width=12.5cm,height=7.5cm]{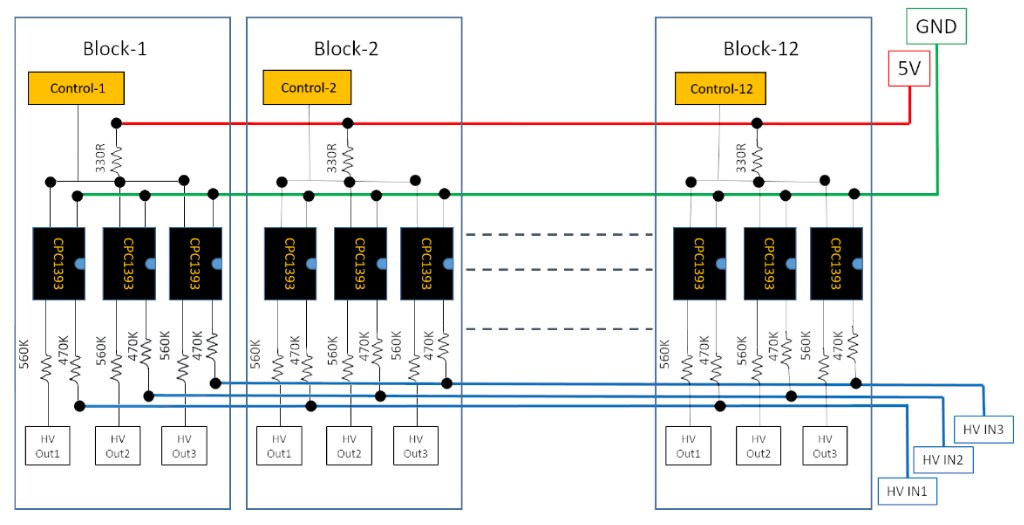}}}
\qquad
\subfloat[]{{\includegraphics[width=7.0cm,height=4.50cm]{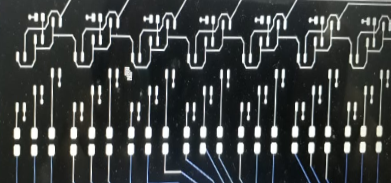}}}
\caption{\label{fig:optocoupler_scheme} \textbf{(a)}: Circuit diagram of the optocoupler scheme. \textbf{(b)}: Image of the drift PCB with provision for staggered optocoupler configuration.}
\end{figure}

The external bias voltage to the foil segments was provided using two ceramic resistive dividers, each biasing one-half of the detector \cite{Kumar:2020plo}. The connection to each segment was regulated using a series of optocouplers, which serve as switches as depicted in Fig.~\ref{fig:optocoupler_scheme} (a). After several design iterations, the optocouplers were ultimately arranged in a staggered configuration on the drift board, as shown in Fig.~\ref{fig:optocoupler_scheme} (b). In the unlikely event of irreversible damage to a particular foil segment, these optocouplers allow us to isolate that region completely, allowing the module to continue functioning with reduced acceptance. Although the drift board's high voltage layout is designed to control this operation remotely using microcontrollers, this task is carried out manually via switches for this experiment. Fig.\ref{fig:resistive_scheme}(c) shows a divider circuit containing various resistors used across drift, transfer, and induction gaps. This circuit is commonly known as a "branch," and the current flowing through it will subsequently be called the "branch current."

Four Common ReadOut Boards (CROBs) \cite{CBM-MuCh:2023gdx} were used to collect data from the FEBs and transmit it to the Common Readout Interface (CRI) \cite{Zabolotny:2017hwz} through high-speed optical fiber cables coupled with GigaBit Transceiver (GBT) links. The CRI then facilitates the data storage in a dedicated storage in binary format. The data is stored in different files that are systematically unpacked and utilized during offline analysis. This stored data could also be used for on-site monitoring to check for potential issues during data acquisition.

While the FLES (Front-Level Event Selector) Interface Boards (FLIBs) were outside the beam cave, the LVDBs and CROBs were mounted on the detector frame and exposed to radiation.

\begin{figure}[htbp]
\centering 
\includegraphics[width=11.0cm,height=5.0cm]{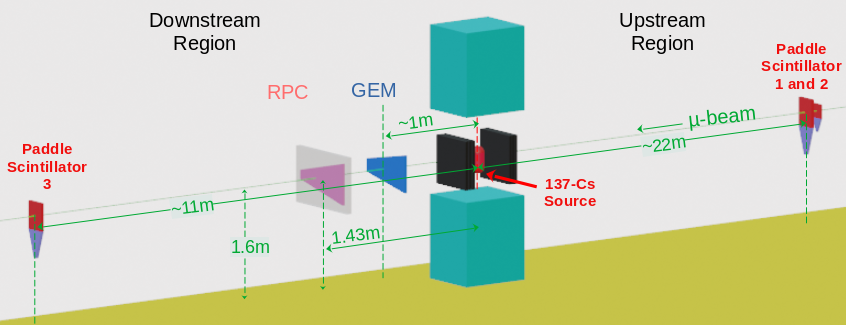}
\caption{\label{fig:SchematicSetupGIF} Schematic setup of the GIF++ beam-test. One GEM and one RPC chamber are placed along the beamline at '1 m' and '1.43 m' from the 137-Cs source, respectively.}
\end{figure}

Fig.\ref{fig:SchematicSetupGIF} shows a sketch of the experimental setup used to perform the detector tests at GIF++. The GEM and RPC are located downstream from the gamma source at approximately '1 m' and '1.43 m', respectively. These detector modules, coupled to the LVDBs and CROBs, were affixed to an aluminum mounting frame. A trapezoidal aluminum plate was attached to the GEM module, facilitating thermal cooling for the FEBs through internal pipes and allowing cold water flow. However, we opted for ambient air cooling for all the FEBs this time.

For GEM operation, a gas mixture of argon and carbon dioxide was used in a 70: 30 volume ratio, with a recorded flow rate of 3.5$\pm$0.25 liters per hour. In contrast, the RPC utilized a gas mixture consisting of R134a, iC$_{4}$H$_{10}$ and SF$_{6}$ in proportions of 95.2\%, 4.5\%, and 0.3\%, respectively, with a humidity level of 40\%.

The source intensity was adjusted using remotely controlled lead (Pb) attenuators. Two large paddle plastic scintillators ('30 cm' x '40 cm' each) and one smaller plastic scintillator ('4.5 cm' x '4.5 cm') placed outside the beam hall formed part of the setup to differentiate muons from the gamma background. The output of a three-fold coincidence (all three paddle scintillators) and a four-fold coincidence (three-fold + RPC) have been used as triggers for our experiment. The former trigger was acquired after feeding the output of the coincidence logic unit to a FEB, while the latter was applied during the offline analysis.

The data collection process involved varying the bias voltage of GEM and the source intensity. We maintained a fixed charge threshold of '6 fC' during calibration for all FEBs. Detailed characterization studies were then performed to evaluate detector efficiency variation, gain fluctuation, time resolution, and rate capability as functions of applied voltage and source intensity. In addition, we moved the detectors to expose different pad areas and switched the RPC and GEM positions to expose each chamber unhindered. The data acquired in the experiment was studied offline using CBMROOT \cite{CBMROOT}. The analysis is divided primarily into two major categories: one with a muon beam in the absence of a gamma source and the other in the presence of a gamma source. 

\section{Results}
\label{sec:data_taking_and_test_results}

\begin{figure}[htbp]
\centering 
\includegraphics[width=8.0cm,height=5.50cm]{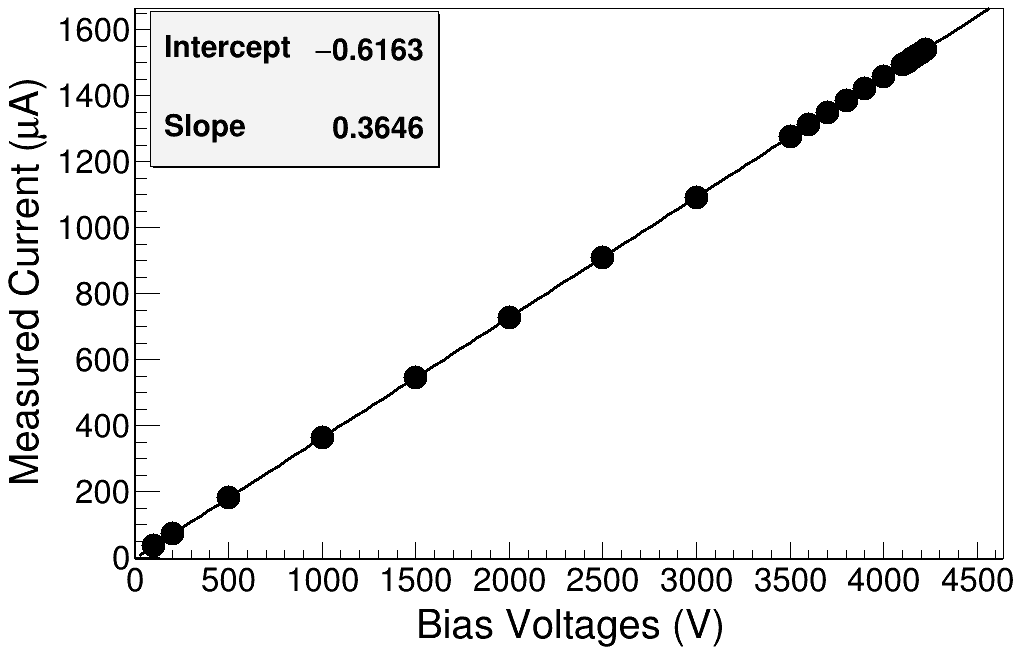}
\caption{\label{fig:voltage_response} Variation of the branch current with applied bias voltage showing a linear trend. Fit function: Intercept + Slope*x}
\end{figure}

Before the beam tests, we performed some routine quality checks: 
\begin{enumerate}
\item \textbf{Voltage scan}: We varied the voltage of the GEM detector and measured the current in the divider chain as we went up toward the upper operational limit. The corresponding plot is shown in Fig.\ref{fig:voltage_response}, illustrating an almost proportional response ('0.364 $\mu$A/V').

% \item \textbf{Gas flow check}: Observations revealed fluctuations in the gas flow rate around the mean, averaging 0.25 liters per hour during the data collection period.
% \item \textbf{Detector stability test}: Operating the detector for an extended duration (1 hour) revealed neither sudden drops in the current flowing through the GEM voltage-divider chain (also known as "branch current") nor discharges, indicating a stable detector ready to acquire data.
\item \textbf{FEB Calibration}: Front-End Boards (FEBs) play a crucial role in collecting analog charge signals from the detector and converting their amplitudes into digital values. The signal shaping time is set to '240 ns.' Each 18 FEB is equipped with tunable 5-bit Analog-to-Digital Converters (ADCs) and was pre-calibrated to ensure a uniform charge response across all channels.

To achieve this, the VRef parameters of all ASICs (VRefP, VRefN, and VRefT) \cite{CBM-MuCh:2023gdx} were adjusted to maintain a uniform threshold of '6 fC' and a bin size of '2.5 fC' per ADC channel. Since each ADC consists of a total of $2^{5} - 1 = 31$ channels (One channel is reserved for fast shaper), the charge values are stored in "ADC units" ranging from 1 to 31, corresponding to a charge range from '6 fC' to '83.5 fC' in steps of '2.5 fC'. The calibration was performed using an external pulse generator to ensure consistency across all channels.
\end{enumerate}

\subsection{Muon beam only}
\label{subsec:analysis_with_muon_only}
For the data with a muon beam only, the focus is on exploring the detector performance by solely varying the bias voltages. The bias voltages have been varied to ensure a steady difference of '10 $\mu$A' in consecutive branch current values. However, we have used the summation of the potential differences across the 3 GEM Foils ($\Delta$V$_{\text{3 GEM}}$) instead of the bias voltage to provide uniformity with conventional texts. Meanwhile, other parameters, such as acceptance, gas flow rate, and threshold, remain unchanged. Each data "run" for this analysis is about 5 minutes long. We have discussed the characteristic results of this analysis in subsequent subsections.

\subsubsection{Noise characteristics}

\begin{figure}[htbp]
\centering 
\subfloat[]{{\includegraphics[width=6cm, height=4cm]{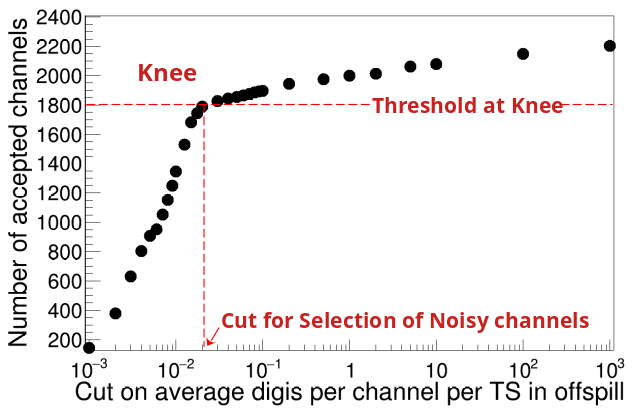}}}
\qquad
\subfloat[]{{\includegraphics[width=6cm, height=4cm]{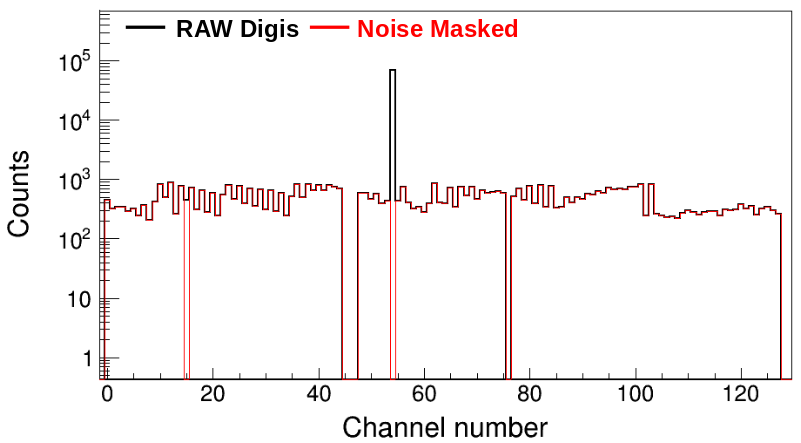}}}
\caption{\label{fig:noise_threshold_plot} \textbf{(a)}: Number of channels (out of 2231) that remain after applying cut on the average number of digis per channel per TS. \textbf{(b)}: Spill structure for muon-only run at HV $=$ 4101 V (without trigger) before noise masking and after. }
\end{figure}

The digi distribution generally comprises signal and electronic noise, with the latter appearing in two distinct forms. First, specific electronic channels exhibit significantly higher data counts than others, known as "hot channels." Second, some channels register elevated counts relative to the average but remain lower than hot channels, referred to as "noisy channels." Hot channels are readily identified and excluded through offline masking. We determine the average number of digis per timeslice (TS, ~'128 ms') for each channel during beam-off periods to detect noisy channels. Ideally, good channels should show negligible digis in this region, but noisy channels exhibit digi counts higher than average. We record channels whose average digis per TS in the off-spill region exceed a specified threshold. Applying this method across a range of thresholds, we generate the characteristic plot shown in Fig.\ref{fig:noise_threshold_plot} (a). The plot shows that as the digi count threshold decreases, the number of noisy channels decreases until a sharp slope change occurs. This inflection point (or "knee") is the cut-off point to distinguish noisy channels from good ones.

\begin{figure}[htbp]
\centering 
\includegraphics[width=7.5cm, height=5cm]{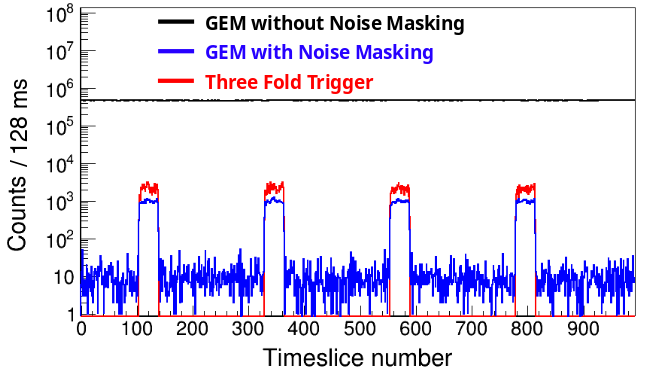}
\caption{\label{fig:SpillStructureAndChannelDist} Typical spill structure as obtained from the digis due to muon beam on GEM detector for the case with no removal of the noisy channels (black) and noise subtracted data (blue). }
\end{figure}

Fig.\ref{fig:SpillStructureAndChannelDist} illustrates the cumulative distribution of digis (also known as spillstructure) from the GEM detector before and after removal of electronic noise (beyond the knee threshold), alongside a 3-fold trigger for reference. According to the distribution, the on-spill region (with beam ON) lasts '4.6 s', while the off-spill region (with beam OFF) extends '23.8 s.' The spill structure after removal of the noisy channels shows a clear overlap of the spills with the 3-fold trigger, demonstrating the importance of noise estimation.

The reduction in noise levels, approaching approximately ten digis per timeslice, as observed in Fig.\ref{fig:SpillStructureAndChannelDist}, demonstrates the efficacy of this electronic noise removal criterion. Fig.\ref{fig:noise_threshold_plot} (b) highlights the effect of the algorithm on the channels of a single FEB. The plot provides insight into the impact of the noise-masking algorithm on the data.

Similar efforts were undertaken for all runs (utilizing muon beam only), compiling a comprehensive list of noisy channels. These identified noisy channels were subsequently eliminated from all runs to ensure uniform spatial acceptance across the board.

\subsubsection{Time resolution, Digi distributions, cluster size, gain and efficiency studies with muon beam}
\label{subsubsec:cluster_size}
\begin{figure}[htbp]
\centering 
\subfloat[]{{\includegraphics[width=6cm,height=4.8cm]{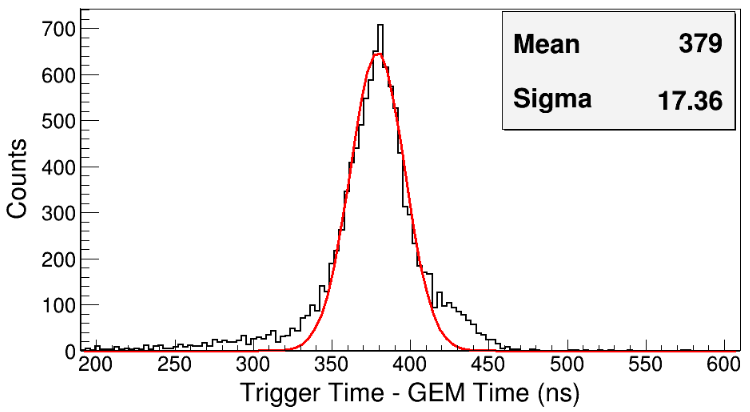}}}
\qquad
\subfloat[]{{\includegraphics[width=8cm, height=4.8cm]{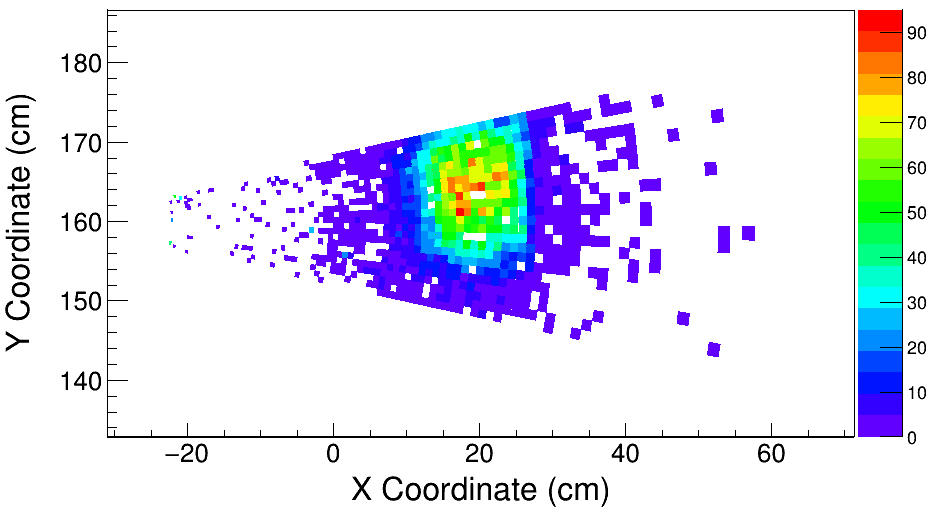}}}
\hspace{0.2cm}
\caption{\label{fig:MuonBeamSpot} \textbf{(a)}: Time-difference spectrum for three-fold scintillator and GEM at $\Delta$V$_{\text{3 GEM}}$ $=$ 1120.9. \textbf{(b)}: Muon beam spot with three-fold trigger after masking hot channels only.}
\end{figure}

In the free-streaming data from our experiment, we expect the detector digis due to muon to be temporally correlated with the coincidence trigger(s). The typical distribution of the time difference between the three-fold trigger and noise-masked GEM digis for a single run (also known as a time-difference spectrum) is depicted in Fig.~\ref{fig:MuonBeamSpot} (a) while Fig.\ref{fig:MuonBeamSpot} (b) shows a 2D distribution of the digis on the GEM plane utilizing the three-fold scintillator output as a trigger and in a 3$\sigma$ time window resulting in a distinct "beam-spot." The peak of the time-difference spectrum is fitted using a Gaussian distribution, where the mean corresponds to the offset between the time stamps registered in FEBs from the particle crossing the GEM detector and the 3-fold scintillator pulse. The standard deviation ($\sigma$) provides insight into the detector time resolution for a given bias voltage. In this figure, we also notice a shoulder on the right and a tail on the left. The former possibly corresponds to delta electrons, which deposit relatively higher energy in the gas gap. The latter is due to the "time-walk" effect, which means small signals cross a fixed threshold later than bigger ones, even synchronous ones.

The mean of the time-difference spectrum remained consistent across all runs at the same bias voltage. In the initial phase of tests at mCBM \cite{mCBM}, it was observed that a sudden change in peak position in the event of a discharge occurs. No such shift in the peak position was observed for the entire GIF++ beamtime, thus highlighting the stability of our detector.

\begin{figure}[htbp]
\centering 
\subfloat[]{{\includegraphics[width=4.5cm,height=4.0cm]{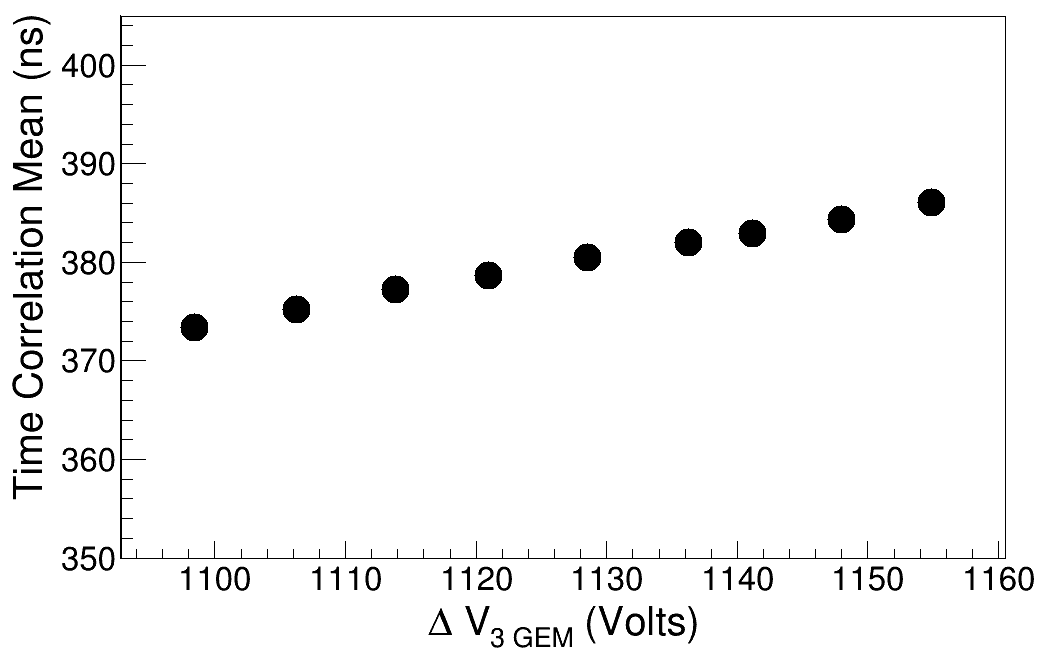}}}
\qquad
\subfloat[]{{\includegraphics[width=4.5cm,height=4.0cm]{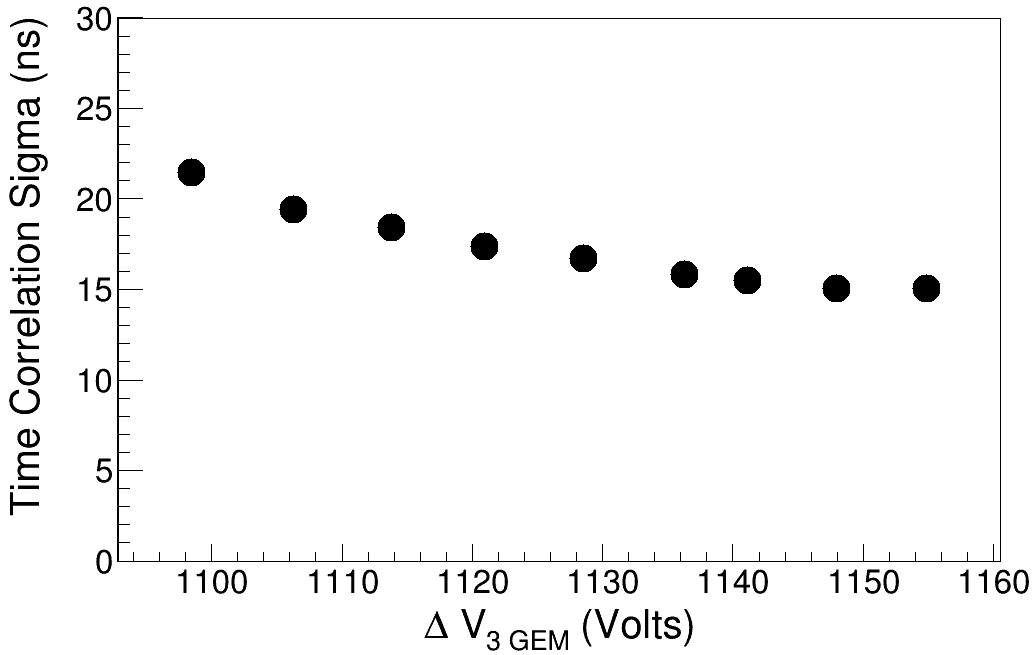}}}
\qquad
\subfloat[]{{\includegraphics[width=4.5cm, height=4cm]
{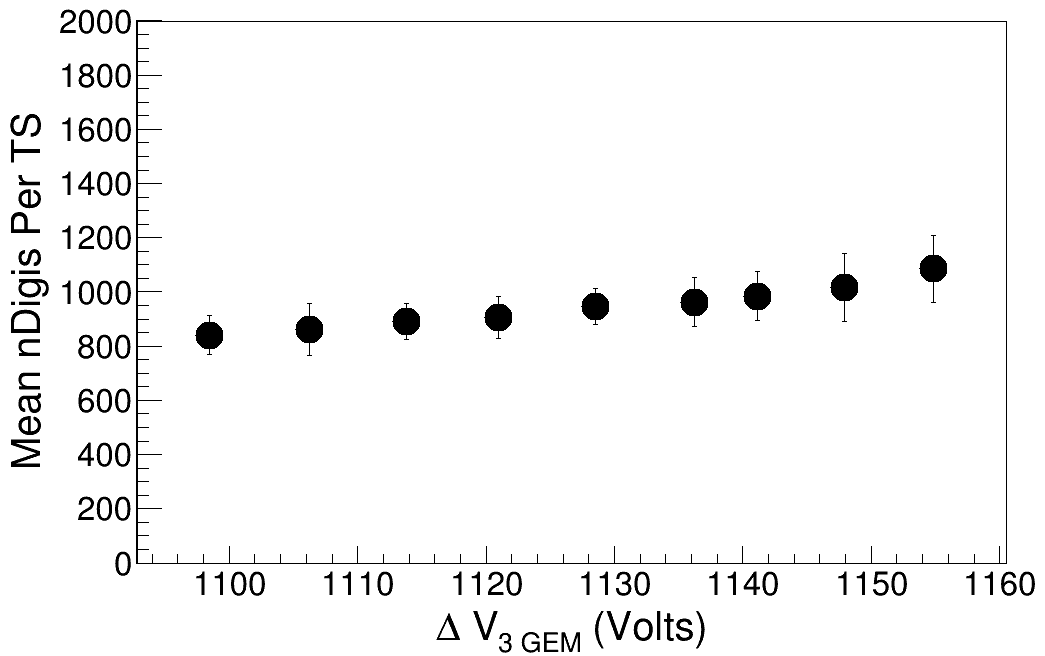}}}
\caption{\label{fig:time_correlation_muon_only_plots} \textbf{(a)}: Time-correlation mean (ns) vs. summed up $\Delta$V$_{\text{3 GEM}}$ (V). \textbf{(b)}: Time-correlation sigma (ns) vs. summed up $\Delta$V$_{\text{3 GEM}}$ (V). \textbf{(c)}: Mean of the number of digis per timeslice (TS) vs. summed up $\Delta$V$_{\text{3 GEM}}$ (V).}
\end{figure}

The dependence of the mean of the time-difference spectra (with statistical error $\sigma$/$\sqrt{N}$) on the GEM voltages summed across the three GEM foils ($\Delta$V$_{\text{3 GEM}}$) is shown in Fig.\ref{fig:time_correlation_muon_only_plots} (a). It is evident that with an increase in bias voltage, the mean of the distribution shifts monotonically. This shift, which reaches up to approximately '15 ns' at the highest voltages, arises due to a decrease in the charge drift time. Furthermore, the time resolution (sigma of the distribution with least-squares fit error) decreases up to a point and then saturates with increasing voltage, as observed in Fig.\ref{fig:time_correlation_muon_only_plots} (b). Fig.\ref{fig:time_correlation_muon_only_plots} (c) shows the variation of the average number of digis per timeslice (nDigis/TS) with the bias voltage. The error bars in the figure represent the statistical errors. With an increase in $\Delta$V$_{\text{3 GEM}}$, we observe a rise in the mean nDigis/TS. This change is attributed to an increase in the mean cluster size with the bias voltage.
Using a time-difference window of $\mu \pm 3\sigma$, we employed a clustering algorithm to determine the total charge collected from an incoming muon candidate and estimated the gain. This algorithm calculates the total charge collected within a 3$\sigma$ time window of the trigger while considering a maximum separation of two pads from the first digi of the cluster in any direction, including vertical, horizontal, or diagonal. The total charge accumulated within a spatial cluster is termed the "cluster charge." It is computed using a charge threshold of '6 fC' and a bin size of '2.5 fC'. The conversion formula \cite{CBM-MuCh:2023gdx} is given by:

\begin{equation} \text{Cluster Charge (fC)} = 6 \text{ (fC)} \times \sum \text{Number of digis} + 2.5 \text{ (fC)} \times \sum \text{(Cluster charge of digis in ADC units)} \end{equation}

\begin{figure}
\centering
\subfloat[]{{\includegraphics[width=4.5cm,height=3.5cm]{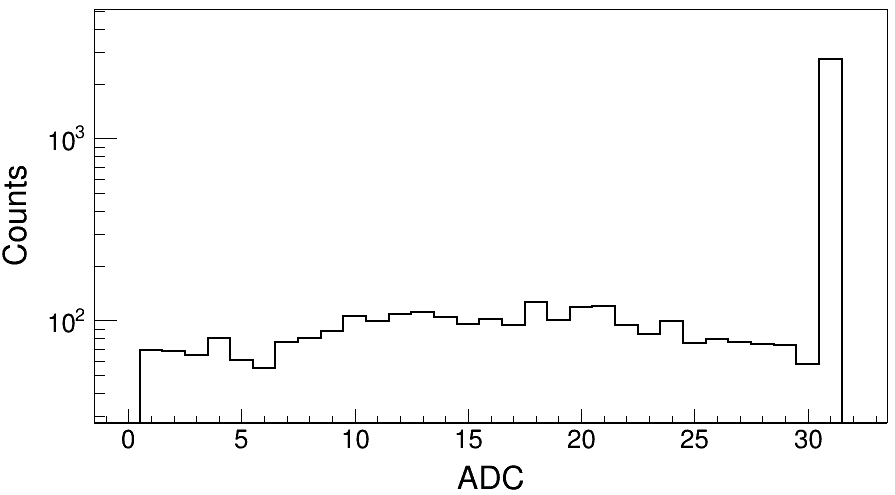}}}
\qquad
\subfloat[]{{\includegraphics[width=4.5cm,height=3.5cm]{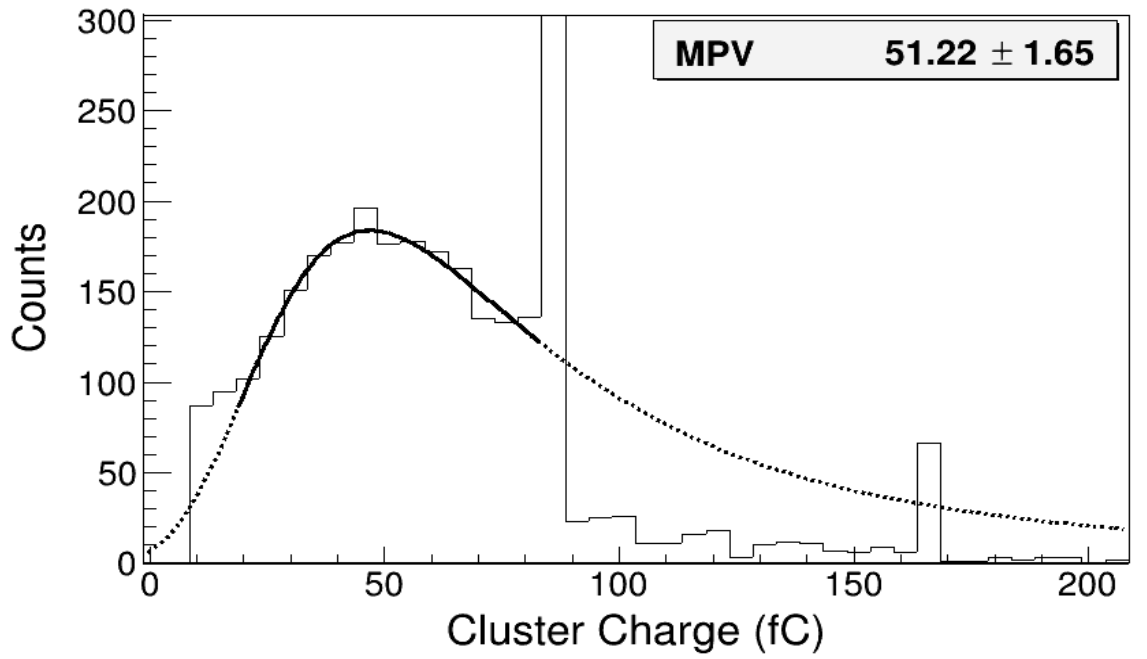}}}
\qquad
\subfloat[]{{\includegraphics[width=4.5cm,height=3.5cm]{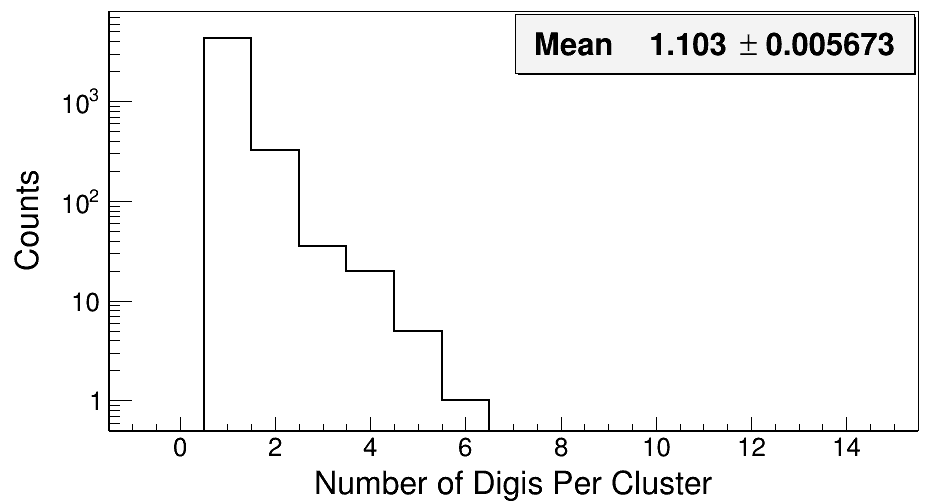}}}
\caption{\label{fig:cluster_adc_muon_only} \textbf{(a)}: Raw charge distribution plot of muon beam digis in ADC units for $\Delta$V$_{\text{3 GEM}}$ $=$ 1120.9 V. 
\textbf{(b)}: Cluster charge distribution for muons at $\Delta$V$_{\text{3 GEM}}$ $=$ 1120.9 V fitted with a Landau distribution.
\textbf{(c)}: Number of digis per cluster distribution for muons at $\Delta$V$_{\text{3 GEM}}$ $=$ 1120.9 V.}
\end{figure}

Fig.\ref{fig:cluster_adc_muon_only}(a) displays the distribution of raw charge collected from all FEBs in ADC units for $\Delta$ V$_{\text{3 GEM}}$ $=$ 1120.9 V, and Fig.\ref{fig:cluster_adc_muon_only}(b) shows the corresponding cluster charge distribution in femtocoulombs. We used a Landau function to fit the distribution. The Most Probable Value (MPV) has been extracted using a fitting range from '18.5 fC' to '83.5 fC' thus avoiding the overflow bin and beyond (indicated by a bold line). A least-squares fit method was used to estimate the error in fit for the MPV values. The spike at '83.5 fC' arises due to the overflow bin corresponding to the last channel of the ADC (31 ADC units). The distribution of the number of digis per cluster with the standard error for muons is shown in Fig.\ref{fig:cluster_adc_muon_only}(c).

\begin{figure}[htbp]
\centering 
\subfloat[]{{\includegraphics[width=4.5cm, height=4cm]
{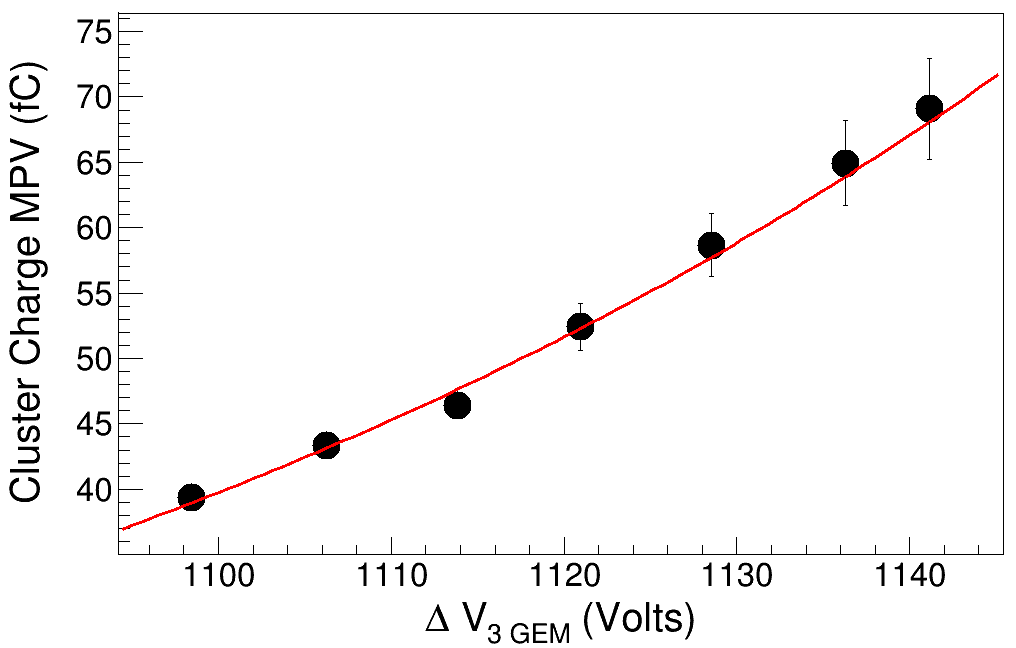}}}
\qquad
\subfloat[]{{\includegraphics[width=4.5cm, height=4cm]
{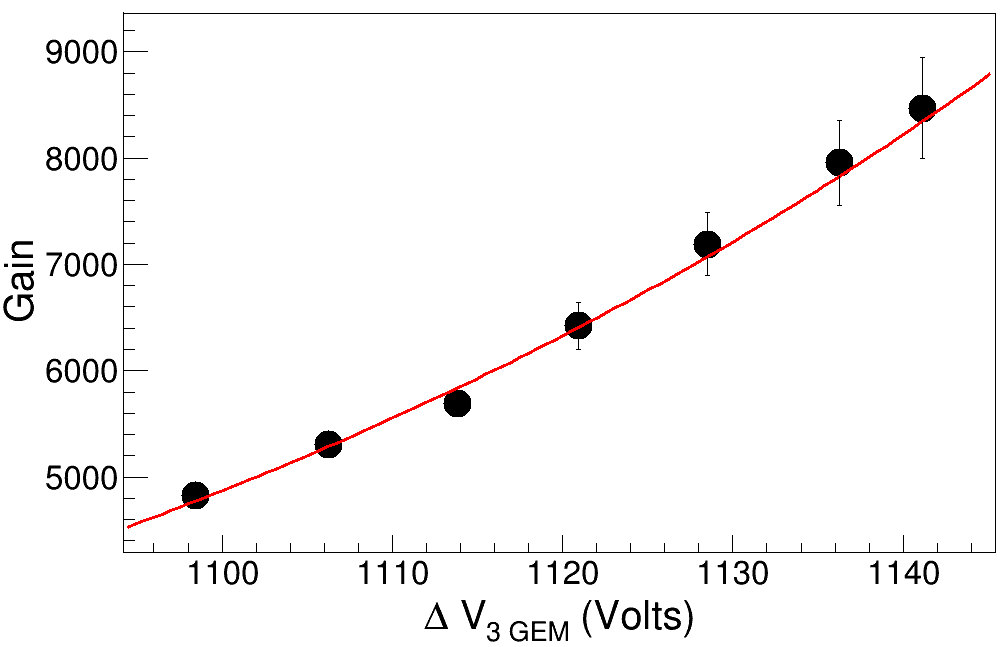}}}
\qquad
\subfloat[]{{\includegraphics[width=4.5cm, height=4cm]
{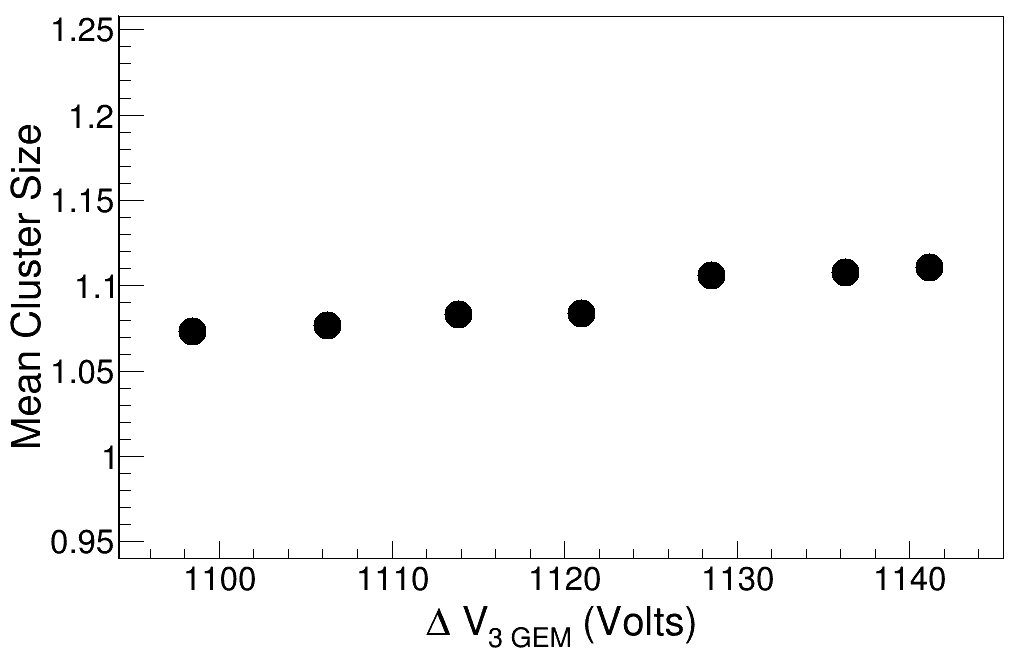}}}
\caption{\label{fig:mpv_vs_voltage} \textbf{(a)}: Cluster Charge MPV vs. summed up $\Delta$V$_{\text{3 GEM}}$ (V) with an Exponential fit. \textbf{(b)}: Effective gain vs. summed up $\Delta$V$_{\text{3 GEM}}$ (V) with an Exponential fit. \textbf{(c)}: Mean cluster size vs. summed up $\Delta$V$_{\text{3 GEM}}$ (V).}
\end{figure}

The variation of the cluster charge distribution (MPV with uncertainty using the least-squares fit method) is plotted with the GEM voltage in Fig.\ref{fig:mpv_vs_voltage} (a). The plot has been fitted with an exponential function, which helps us estimate the effective gain at any voltage by linking the Landau MPV to the number of primary electron-ion pairs produced by muons of momentum (p) $=$150 GeV/c in GIF++. For Minimum Ionizing Particles (MIPs), this number is 100 primary ion pairs per cm \cite{Titov:2012yoy}. The differential energy deposition (dE/dx) of muons in GIF++ is found to be approximately 70\% higher than MIPs using the GEANT4 simulation, giving 51 ion-electron pairs in the active volume region. 

Therefore, the value of effective gain is given by the formula:
\begin{equation}
G_{eff} = \dfrac{\text{M}}{N_{P} \cdot q_{E}}
\end{equation}
where, \\
$G_{eff}$ is the effective gain, \\
M is the MPV value of the cluster charge distribution in femto-Coulombs (fC), \\
$N_{P}$ is the mean number of primary ion pairs created, and \\
$q_{E}$ is the electron charge in fC.

The gain calculated using this formula and its variation with $\Delta$V$_{\text{3 GEM}}$ is shown in Fig.\ref{fig:mpv_vs_voltage} (b). The uncertainty in gain is calculated by the method of propagation of error. At $\Delta$V$_{\text{3 GEM}}$ $=$ 1120.9 V, the mean gain is around 6400. Fig.\ref{fig:mpv_vs_voltage} (c) shows a graph of the change in the mean number of digis per cluster (or cluster size) with standard error as a function of varying voltage. It shows an increase in the mean cluster size with voltage.

%\subsubsection{Detector Efficiency with varying branch current}
Efficiency is another equally important parameter for any detector studies. It is defined as the ratio of the number of detected particles to the number of particles passing through the detector. In general, the efficiency of charged particle detection of GEM can be calculated by taking the number of instances where at least one recorded GEM digi exists in the 3$\sigma$ time window of any 3-fold trigger and dividing it by the total number of 3-fold triggers. Unfortunately, not all muons pass through our detector module's active area (or acceptance). An unknown number of muons travel outside the acceptance or through masked channels. To address this, we used a four-fold trigger system involving three scintillators and two adjacent RPC pads (each of dimension '1.25 cm' $\times$ '1.25 cm') in the overlap zone of GEM. The overlapping region was carefully chosen to ensure no dead channels were present in the overlapping GEM plane. 

\begin{figure}[htbp]
\centering 
\includegraphics[width=7.5cm, height=5cm]{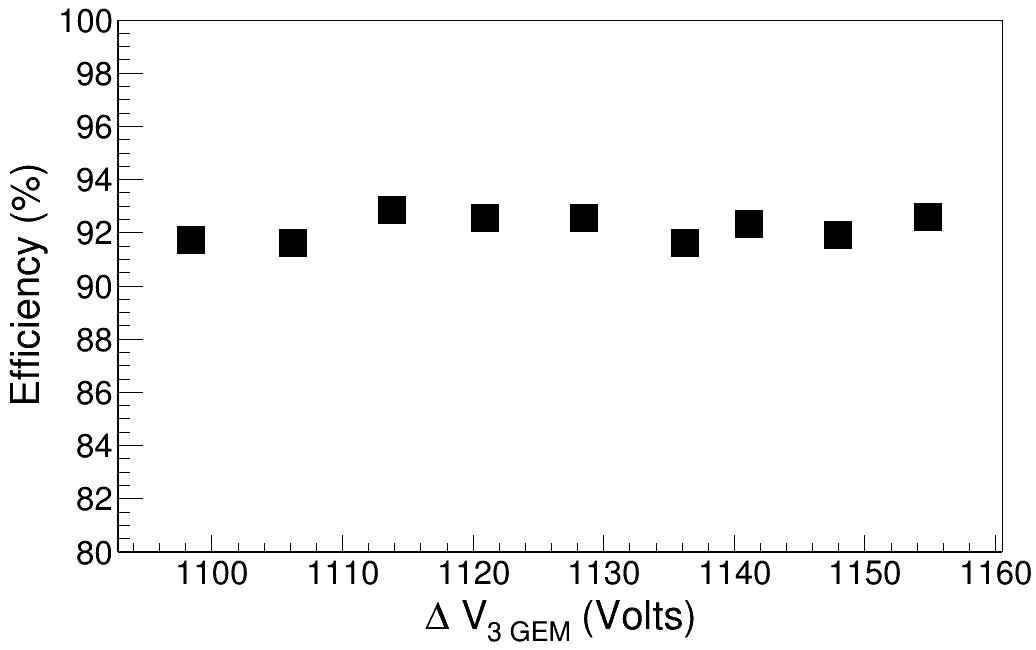}
\caption{\label{fig:EffvsVoltage} 4-fold GEM efficiency vs. summed up $\Delta$V$_{\text{3 GEM}}$ (V) on the GEMs.}
\end{figure}

Fig.\ref{fig:EffvsVoltage} presents the 4-fold efficiency values, indicating an average efficiency of approximately 91.6 $\pm$ 0.2\%. The uncertainty in efficiency is determined using the binomial error calculation, although it remains smaller than the marker size. It is worth noting that the efficiency values obtained through this method are inherently underestimated due to blind regions arising from the gaps in foil segmentation and readout. The reported efficiency should not be interpreted as an absolute measurement without a precise tracking detector. Instead, this analysis demonstrates the stability of the efficiency of the GEM module over a broad range of operating voltages and, by extension, gain values.

\subsection{Muon beam in the presence of external gamma source and varying intensity}
\label{subsec:analysis_with_muon_and_gamma_source}

\begin{figure}[htbp]
\centering 
\subfloat[]{{\includegraphics[width=7cm, height=5cm]{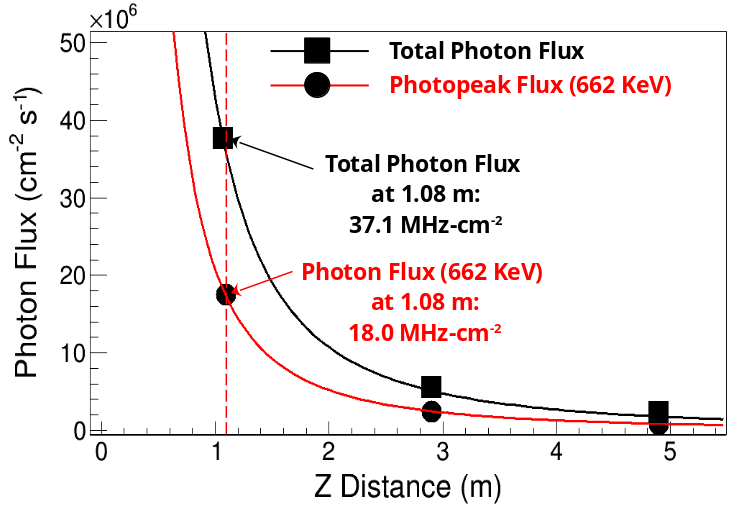}}}
\qquad
\subfloat[]{{\includegraphics[width=7cm, height=5cm]{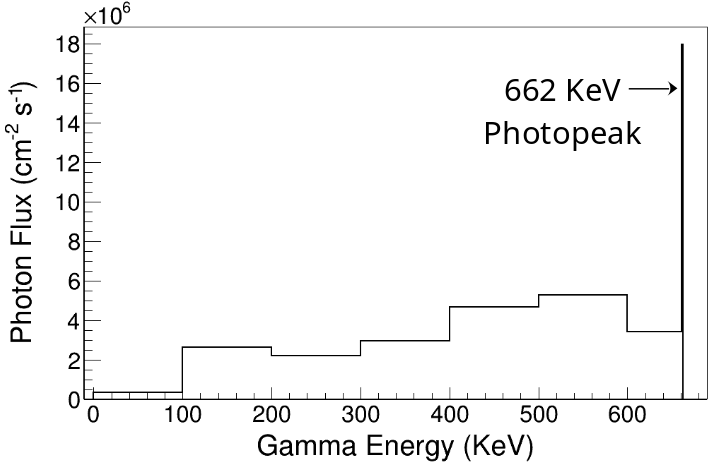}}}
\caption{\label{fig:Photon Flux vs. Z} \textbf{(a)}: Simulated photon flux (Total and Photopeak) for Cs-137 source as a function of $z$ coordinate at the date of the experiment. Black dots and squares represent the simulated flux at that distance, and the function is fitted with 1/$z^{2}$. \textbf{(b)}: Simulated energy-wise gamma flux distribution with variable binning size at $z$=1.08 m}
\end{figure}

Using the same setup and noise characteristics as in Section \ref{subsec:analysis_with_muon_only}, we analyzed the data with a muon beam in the presence of a gamma background. Fig.\ref{fig:Photon Flux vs. Z} (a) presents a simulation plot for the gamma flux inside the GIF++ cave as a function of distance ($z$) from the source (cm), fitted with the 1/$z^{2}$ function. Similarly, Fig.\ref{fig:Photon Flux vs. Z} (b) shows the relative compositions of different gamma energies at a distance of $z$=1.08 m. The simulation results have been taken from Ref.\cite{Pfeiffer:2016hnl} after making corrections for the activity of the gamma source at the time of the experiment. We used the total gamma background instead of only the '662 KeV' photopeak for our analysis since the former provided a more accurate description of the particle flux. 

\begin{figure}[htbp]
\centering 
\includegraphics[width=7.5cm, height=5cm]{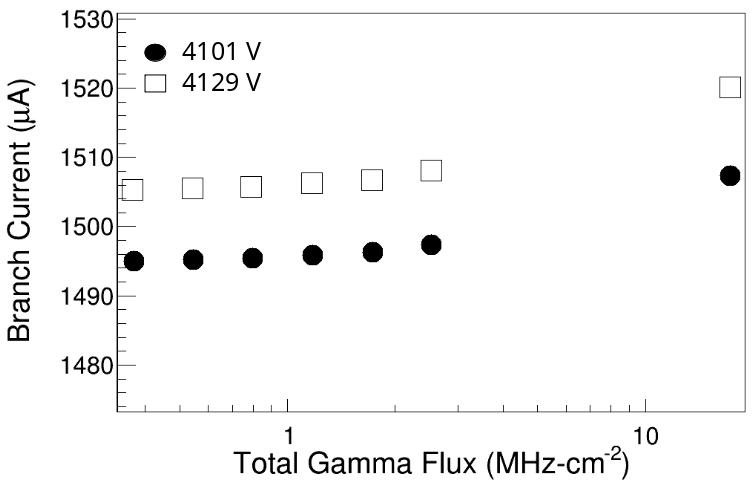}
\caption{\label{fig:BranchCurrentVsIntensity} Branch current vs. total gamma flux for two different bias voltages}
\end{figure}

It is important to note that, unlike Section \ref{subsec:analysis_with_muon_only}, where $\Delta$V${\text{3 GEM}}$ was used, this section utilizes the total bias voltage (HV). This choice is made because the branch current, and consequently $\Delta$V${\text{3 GEM}}$, could vary by a few volts with the intensity of the incident gamma, as illustrated in Fig.\ref{fig:BranchCurrentVsIntensity}, since the high incident flux reverts to a high amplification current, which can cause voltage drops across the resistors placed along the power distribution lines, resulting in a lower potential. This current can cause a voltage drop across the protection resistor, resulting in a lower effective potential across the GEM foil than the actual voltage. Hence, the effective drop across the GEM foils ($\Delta$V$_{\text{3 GEM}}$) cannot be measured by our setup. 

This part of the analysis was conducted at two operating voltages. The first setting, HV = 4101 V, corresponds to $\Delta$V$_{\text{3 GEM}}$ = 1120.9 V with a branch current of 1495 $\mu$A. The second setting, HV = 4129 V, corresponds to $\Delta$V$_{\text{3 GEM}}$ = 1128.5 V with a branch current of 1505 $\mu$A.

After a brief period of operation at the highest gamma intensity, the DAQ system encountered backpressure. To prevent potential data loss, data acquisition at this intensity was divided into multiple runs of 10 seconds each. The recorded runs were later merged, ensuring that there was no data loss.

\subsubsection{Time resolution}

\begin{figure}[htbp]
\centering 
\subfloat[]{{\includegraphics[width=7cm, height=5cm] {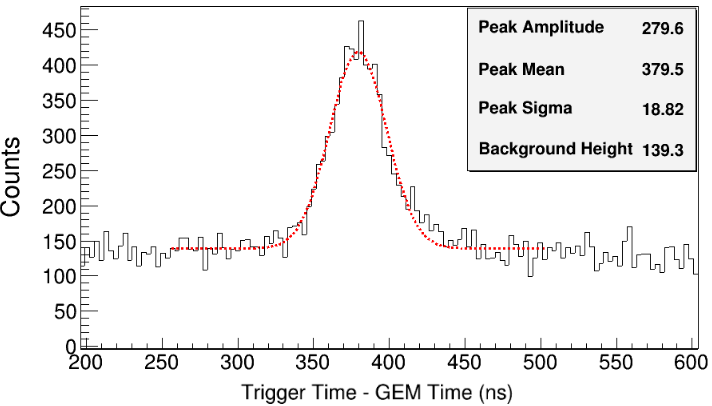}}}
\qquad
\subfloat[]{{\includegraphics[width=7cm, height=5cm]{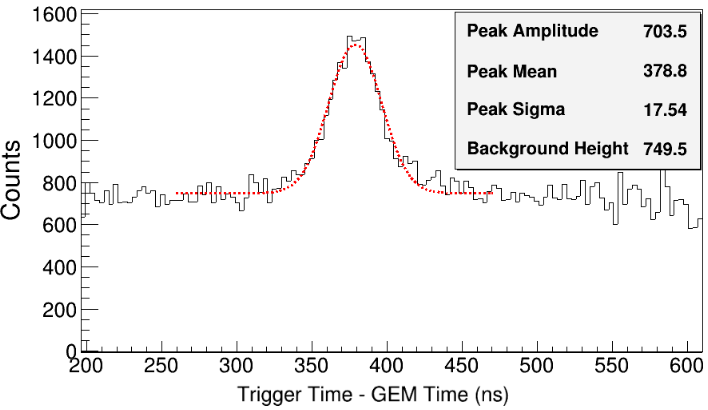}}}
\caption{\label{fig:tcorr_with_intensity_hv_4101 V} \textbf{(a)}: Time-correlation plot with 3-fold trigger for muons with a gamma background of '1.175 MHz/cm$^{2}$' at HV $=$ 4101 V. \textbf{(b)}: Time-correlation plot with 3-fold trigger for muons with a gamma background of '2.52 MHz/cm$^{2}$' at HV $=$ 4101 V.}
\end{figure}

\begin{figure}[htbp]
\centering
\subfloat[]{{\includegraphics[width=7cm, height=5cm]{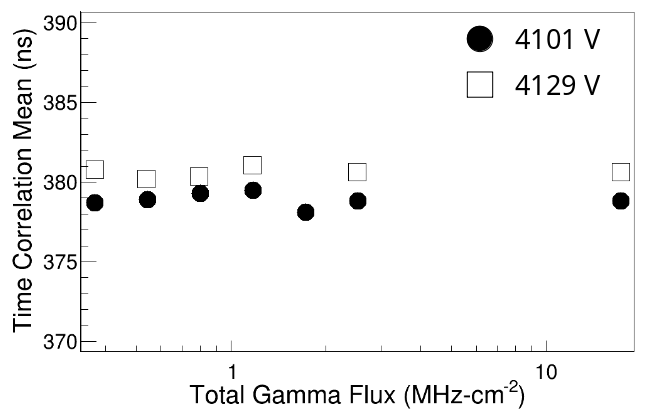}}}
\qquad
\subfloat[]{{\includegraphics[width=7cm, height=5cm]{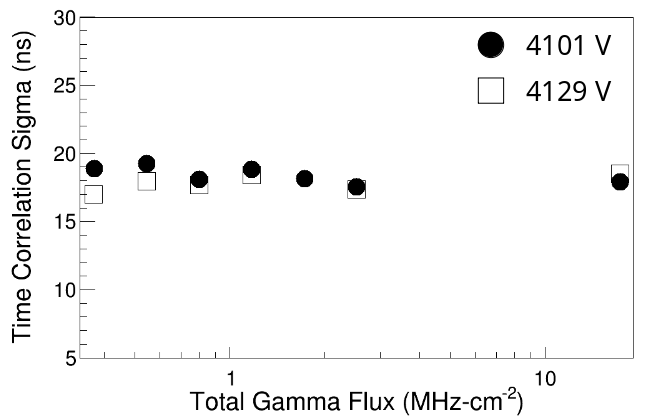}}}
\caption{\label{fig:MeanSigmaInt} \textbf{(a)}: Mean of the time-difference with total gamma flux at two different voltages. \textbf{(b)}: Sigma of the time-difference with total gamma flux.}
\end{figure}

The time-difference spectra for two different values of incident gamma flux are shown in
Fig.\ref{fig:tcorr_with_intensity_hv_4101 V}. A notable difference between these graphs and the previous plots (without a gamma source) is the presence of a background below the Gaussian peak. Applying a method similar to that described in Subsection \ref{subsubsec:cluster_size} for all intensity values results in Fig.\ref{fig:MeanSigmaInt}(a) and (b), which offer a detailed analysis of the mean and sigma of the distributions as a function of incident gamma flux.
We note that the mean and time resolution change with the increase in intensity is either very small or negligible. This observation shows that the detectors can tolerate high gamma flux reliably without a significant shift in signal propagation time or resolution.

\subsubsection{Digi distributions}

\begin{figure}[htbp]
\centering 
\includegraphics[width=9cm, height=5.5cm]{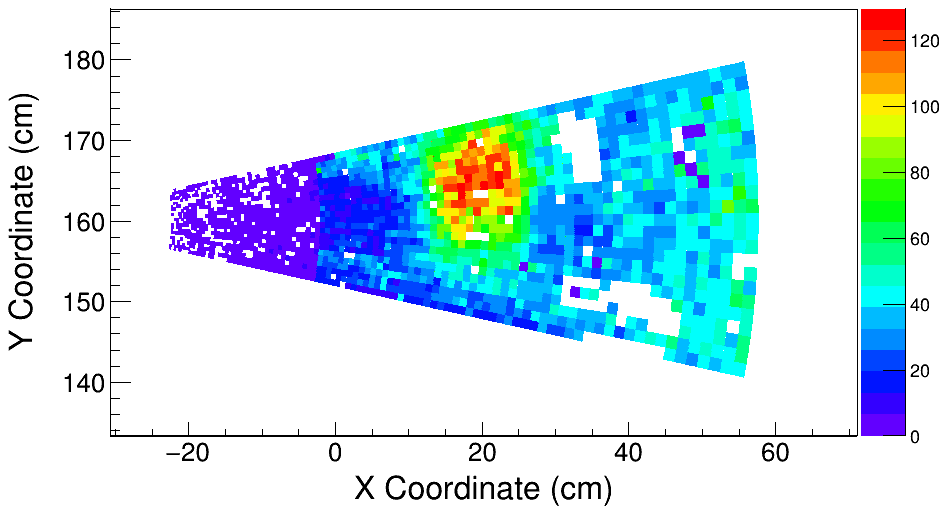}
\caption{\label{fig:BeamSpotWithIntensity} 2D Digi distribution with 3-fold trigger within a 3$\sigma$ time window in the presence of a gamma background of '2.52 MHz/cm$^{2}$' at HV $=$ 4101 V after masking hot channels only.}
\end{figure}

Fig.\ref{fig:BeamSpotWithIntensity} shows a 2D digi distribution for the detector in the presence of a gamma source with a three-fold coincidence trigger in a 3$\sigma$ time window. The detector position is identical to the one performed with data without gamma flux during the analysis. A discernible beam spot region (in red) is observed. Unlike the case without a gamma background, we can see that there are digis even in the region far away from the beam spot, further proving the existence of an "in-time" gamma background. The fact that our detector can discern the beam spot in such a high-intensity gamma environment shows the robustness of our analysis methodology. A sudden drop in occupancy is observed near $X =$ 0. This shift is because the GIF++ source has a limited numerical aperture, which is rectangular. Consequently, our detector's narrow region extends into the gamma source's shadow region. However, since this region is far from the beam spot, it would not affect the results discussed in this paper.

\subsubsection{Gain stability}
\label{subsec:detector_gain}

\begin{figure}[htbp]
\centering 
\subfloat[]{{\includegraphics[width=7 cm, height=5.0cm]{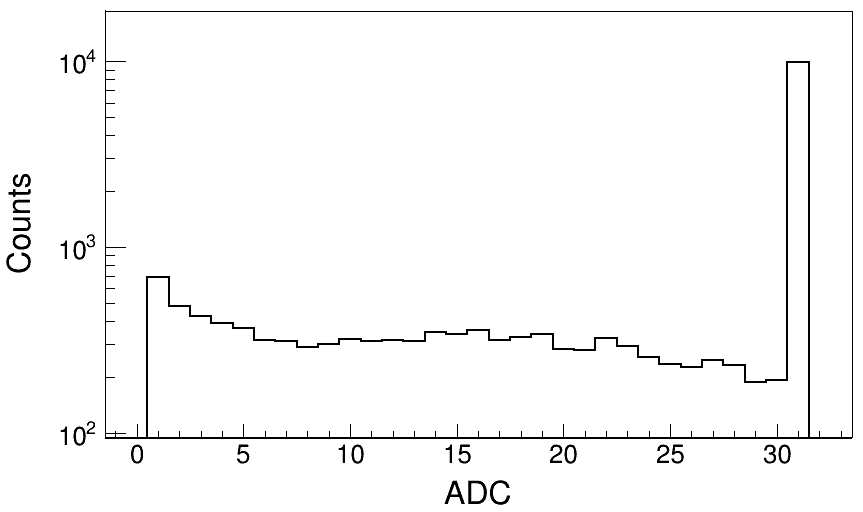}}}
\qquad
\subfloat[]{{\includegraphics[width=7 cm, height=5.0cm]{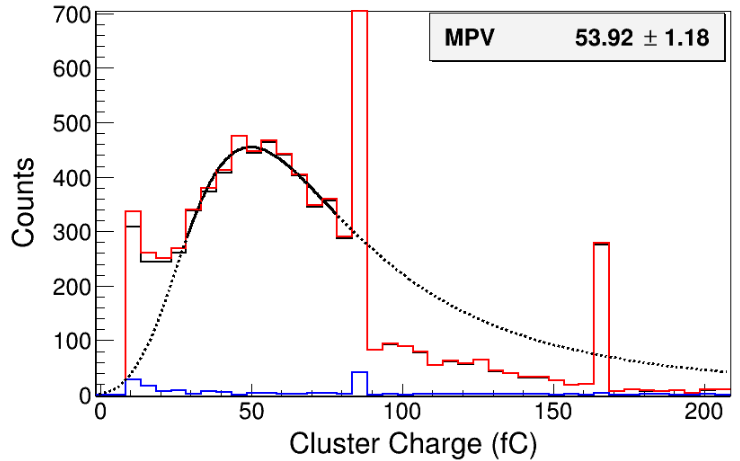}}}
\caption{\label{fig:cluster_charge_with_intensity_hv_4101V} \textbf{(a)}: Raw charge plot for GEM in ADC units with a gamma background of '0.55 MHz/cm$^{2}$' for HV $=$ 4101 V. \textbf{(b)}: Raw (Red), Background (Blue) and Corrected (Black) cluster charge plot for muons with a gamma background of '0.55 MHz/cm$^{2}$' for HV $=$ 4101 V. The corrected distribution is fitted with a Landau function, excluding the overflow bin.}
\end{figure}

Following the methodology outlined in Section \ref{subsubsec:cluster_size}, we investigated the cluster charge distributions across all intensity values at two high-voltage settings: HV = 4101 V and HV = 4129 V. The raw charge distribution (in ADC units) for HV = 4101 V at a gamma intensity of '0.55 MHz/cm²' is presented in Fig. \ref{fig:cluster_charge_with_intensity_hv_4101V}(a). The cluster charge distribution derived from the raw charge data, as shown in Fig. \ref{fig:cluster_charge_with_intensity_hv_4101V}(b) (red), represents a combination of signal and gamma-induced background. To minimize background contributions on the MPV, we have tried to identify the gamma contribution separately by using data from the off-spill region.

This was accomplished using the same time window and a clustering algorithm to correlate the triggers from the on-spill region with a random timeslice from the off-spill region. This criterion is based on the expectation that the off-spill region will have only gamma-induced clusters. The gamma contribution, extracted using this approach and depicted in Fig. \ref{fig:cluster_charge_with_intensity_hv_4101V}(b) (blue), was subsequently subtracted from the raw distribution to obtain the corrected cluster charge distribution, shown in Fig. \ref{fig:cluster_charge_with_intensity_hv_4101V}(b) (black). The corrected plot was fitted with a Landau function in the same range ('18.5 fC' to '83.5 fC') as discussed for the muon-only case. The error in the fit is obtained using the least-squares fit method.

\begin{figure}[htbp]
\centering 
\subfloat[]{{\includegraphics[width=7cm, height=5.0cm]{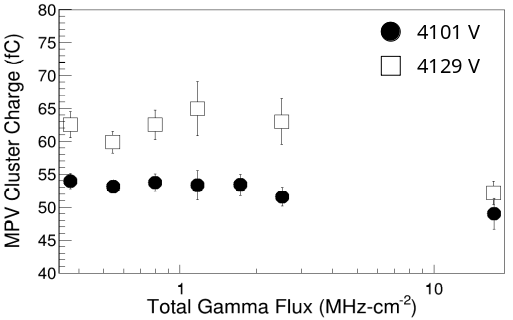}}}
\qquad
\subfloat[]{{\includegraphics[width=7cm, height=5.0cm]{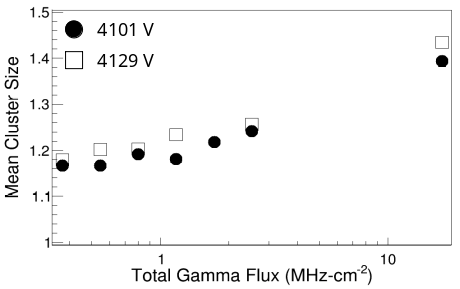}}}
\caption{\label{fig:MPV vs Intensity} \textbf{(a)}: Most probable value (MPV) of the background-subtracted cluster charge distribution vs. total gamma flux for two different voltages. \textbf{(b)}: Mean Cluster Size vs. total gamma flux for two different voltages after background subtraction.}
\end{figure}

Fig.\ref{fig:MPV vs Intensity} (a) shows the MPV values (with least-squares fit error) of the corrected cluster charge distribution as a function of the gamma intensity for two different voltages. One can observe a drop of 10 $\pm$ 5\% for HV $=$ 4101 V, while it falls about 16.5 $\pm$ 3\% for HV $=$ 4129 V at the highest intensity. The error bars are calculated from the fit using the least-squares fit method. The reason for this drop in MPV is assumed to be due to a significant increase in the current (also known as segment current) flowing through the holes in the GEM segments owing to the avalanche created. This effect becomes significant at large particle fluxes, resulting in a voltage drop across the bias resistance of the foil segment, leading to a drop in effective GEM voltage and, hence, the gain (or MPV). As observed in Subsection \ref{subsubsec:cluster_size}, the efficiency values do not change for the MPV range of '40 fC' to '70 fC'. Hence, despite the drop observed in the MPV values (10\% and 16.5 \%), we expect little to no difference in the charged particle detection efficiency.

Fig.\ref{fig:MPV vs Intensity} (b) shows the mean cluster size distribution (with standard error), which increases monotonically with increasing flux. This effect arises from the "random clustering" of low-charge digis from gamma interactions within the trigger window.

While gain values have been tested for stability against fluctuations in electronic parameters such as threshold and shaping time, previous studies \cite{Kumar:2021xpj} have demonstrated that large-area GEM modules often exhibit non-uniformities in the drift gap, which can significantly affect gain calculations. A drift gap variation of 0.5 mm introduces an uncertainty of 16.6\%. Based on variations observed in time resolution measurements from a full-scale module with an identical gap and active area configuration in mCBM, we estimate the drift gap non-uniformity to be approximately 25\%. The statistical uncertainties in the gain values measured at 4101 V and 4129 V are approximately 5\% and 3\%, respectively. Combining these uncertainties in quadrature, we estimate a total uncertainty of less than 30\% for the gain values at these operating voltages. In addition, the gain estimation is limited to a specific area of the detector instead of multiple areas. This means that the local fluctuations of these values could not be estimated.

Furthermore, while our clustering algorithm is designed to minimize the impact of chance coincidences from gamma interactions, it remains sensitive to the chosen time window and spatial parameters. Consequently, this study has not thoroughly examined the effect of clustering on the Most Probable Value (MPV) of the charge distribution. As a result, the absolute gain value derived from the cluster charge MPV has not been precisely determined. Instead, the focus of this exercise is to evaluate the overall stability of the detector in a high gamma-rate environment by analyzing variations in gain as a function of voltage and intensity.

\section{Rate capability of the GEM detector}
\label{subsec:rate_handling_capability_of_the_detector}

\begin{figure}[htbp]
\centering 
\subfloat[]{{\includegraphics[width=7.0cm,height=5.0cm]{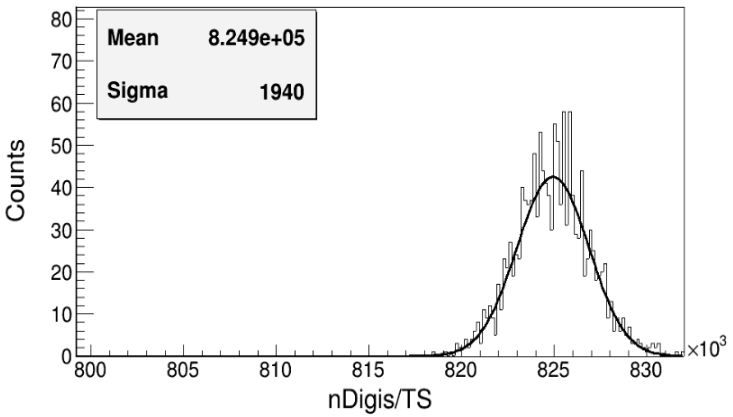}}}
\qquad
\subfloat[]{{\includegraphics[width=7.0cm,height=5.0cm]{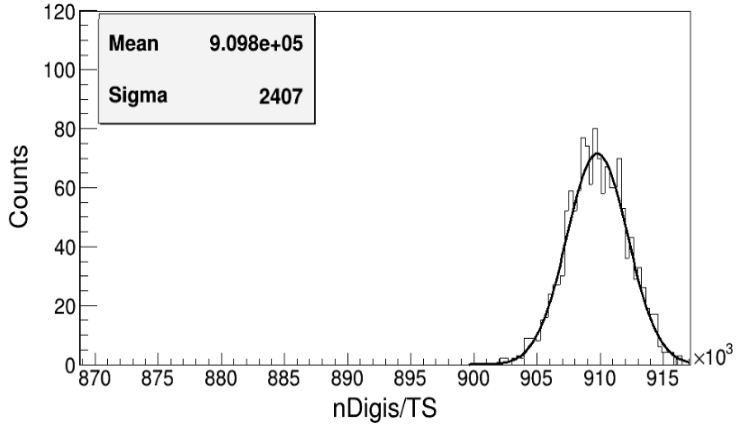}}}
\caption{\label{fig:ndigis_per_TS_plots} number of digis per TS plot for an intensity of 1.175 MHz/cm$^{2}$ at \textbf{(a)}: HV $=$ 4101 V \textbf{(b)}: HV $=$ 4129 V.}
\end{figure}

The study of the rate capability of a detector is crucial in high-rate environments. Determining whether the detector module can withstand extreme radiation conditions without significantly degrading its performance is important. We estimate the data rate corresponding to the high-intensity flux by measuring the digi rate observed by the detector. To understand the digi rate at a given intensity, we plot a 1D histogram of the number of digis per TS and fit it with a Gaussian. Fig.\ref{fig:ndigis_per_TS_plots} illustrates the fit for a given incident gamma intensity for two different bias voltages. 

\begin{figure}[htbp]
\centering 
\subfloat[]{{\includegraphics[width=7.0cm,height=5.5cm]{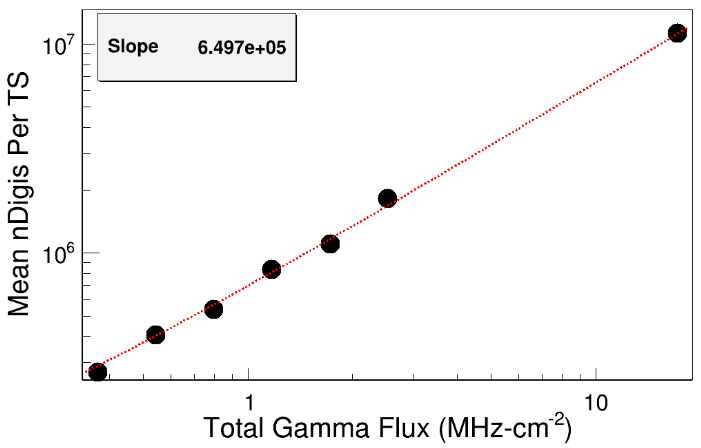}}}
\qquad
\subfloat[]{{\includegraphics[width=7.0cm,height=5.5cm]{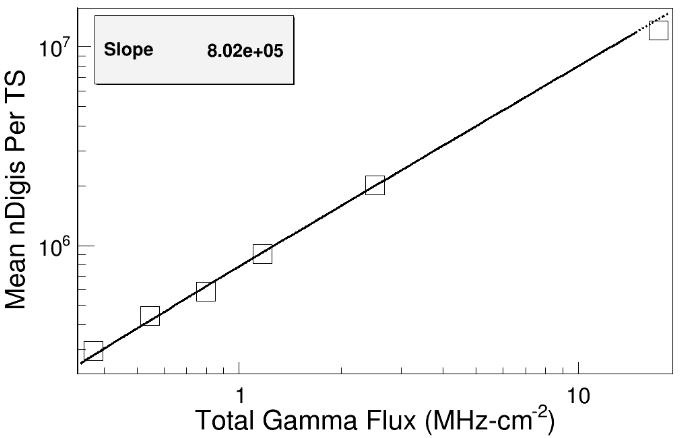}}}
\hspace{0mm}
\vspace{-3mm}
\caption{\label{fig:nDigiswithIntensity} \textbf{(a)}: Average number of digis per timeslice (TS) plot for 4101 V with varying gamma flux. \textbf{(b)}: Average number of digis per timeslice (TS) plot for 4129 V with varying gamma flux.}
\end{figure}

\begin{figure}[htbp]
\centering 
\subfloat[]{{\includegraphics[width=6.5cm,height=5.5cm]{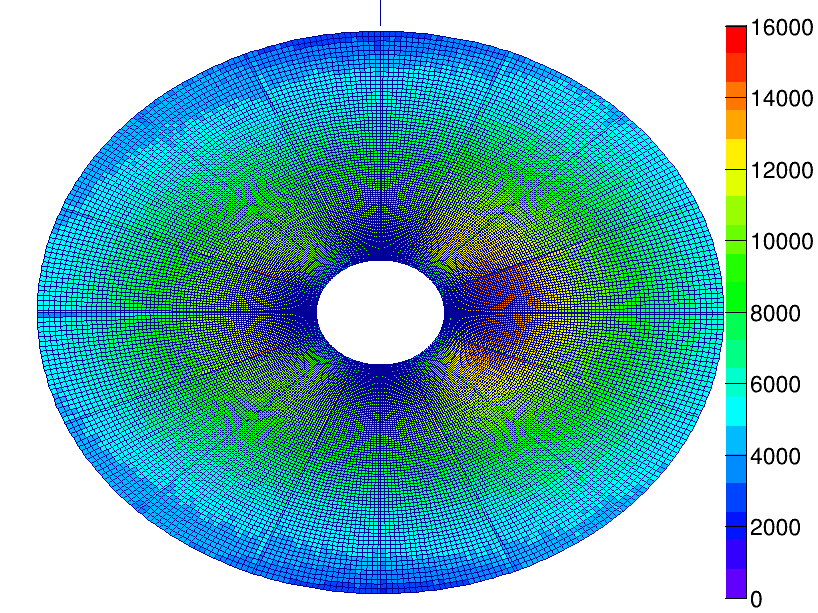}}}
\qquad
\subfloat[]{{\includegraphics[width=7.0cm,height=5.0cm]{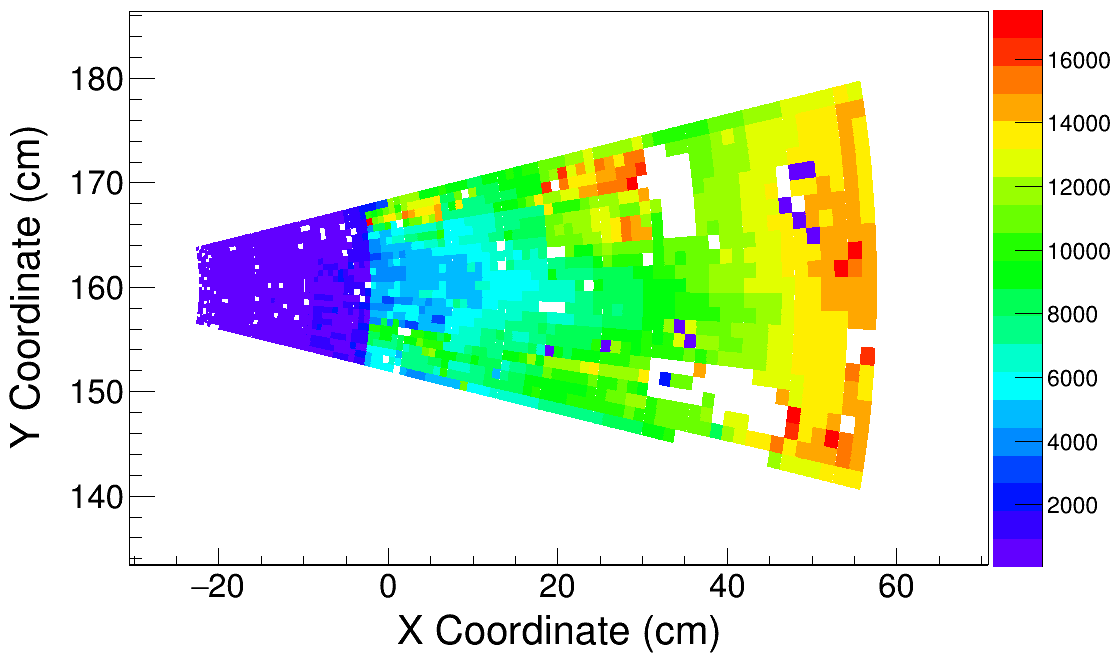}}}
\hspace{0mm}
\caption{\label{fig:simulation_results} \textbf{(a)}: Simulation of 2D digi distribution of all 16 modules of the first layer of station 1 of CBM muon chambers for a duration of 1 TS at the maximum planned collision rates. \textbf{(b)}: 2D digi distribution per TS of GIF++ module at HV $=$ 4101 V without any trigger at highest intensity averaged over 100 Timeslices.}
\end{figure}

After extracting the mean number of digis per TS for all recorded gamma rates at two different voltages, i.e., 4101 V and 4129 V, we study its variation with intensity. A proportional behavior between digi rate and incident gamma flux would indicate a regular response of the GEM chamber. Any non-linearity in response might hint at a lack of formation of digis due to a reduction in gain, efficiency, or both. The plot in Fig.\ref{fig:nDigiswithIntensity} (a) shows a linear response of average digis with intensity for HV $=$ 4101 V. The error bars are obtained after calculating the standard error in the Gaussian distribution. This means that no significant loss exists in the data, thereby not affecting the efficiency of the detector. Thus, our detector is expected to handle high gamma rates at this bias voltage. For HV $=$ 4129 V, as shown in Fig.\ref{fig:nDigiswithIntensity} (b), the reduction of counts at the highest intensity point by roughly 12$\%$ is assumed to be due to shifting of the gammas with small charge collection below the threshold charge of '6 fC' as a result of gain drop, as can be understood from Fig.\ref{fig:MPV vs Intensity} (a). However, we do not expect a significant effect on the muon detection efficiency, as its MPV peaks at a much higher charge for this voltage setting.

The first layer of station 1 of MuCh in CBM will observe the highest intensity. The simulated digi rate distribution at this layer for Au-Au collisions at '12 AGeV' and for an interaction rate of '10 MHz' is shown in Fig.\ref{fig:simulation_results}(a). This 2D distribution was obtained from CBMROOT simulations using a realistic geometry. Events worth 1 TS were considered to compare this result with the data under a similar (rate) condition. The digitization parameters, such as a threshold of '6 fC' and a gain of 6400, were also selected to closely resemble the detector settings at HV = 4101 V in GIF++.
  
Fig.\ref{fig:simulation_results}(b) shows the pad-wise digi rate distribution observed from data per TS at '17.25 MHz/cm$^{2}$' gamma flux which reveals the highest achieved digi rate of about '125 KHz/pad'. Thus, this number is comparable to the highest expected digi rates in CBM, which, according to the simulation, is approximately '120 KHz/pad'.

\section{Summary and outlook}
\label{sec:summary_and_outlook}
Large-area triple Gas Electron Multiplier (GEM) detectors will be installed at the first and second stations of the Muon Chambers (MuCh) system in the Compressed Baryonic Matter (CBM) experiment. In preparation, a real-size triple GEM prototype detector for the first station was tested at the upgraded Gamma Irradiation Facility (GIF++) using a muon beam with varying background gamma intensity. The detector was exposed to high gamma flux up to '17.25 MHz/cm$^{2}$'. All detector components (except GEM foils) were manufactured in India. The detector used Ar/CO$_{2}$(70:30) gas mixture. The detector exhibited a stable response without offset fluctuations over prolonged operation, achieving a best time resolution of approximately 15 ns.

At operating voltage ($\Delta$V$_{\text{3 GEM}}$=1120.9 V), the average cluster size was around 1.1 for the muon-only case in a time-difference window of 3$\sigma$ around the mean. In the same time window, at the highest gamma flux, we observe an increase in mean cluster size up to approximately 1.4. The charged particle detection efficiency measured in 4-fold  mode and without any Gamma background was stable for a wide range of voltages at around 91.6 $\pm$ 0.2 $\%$. However, these efficiency values are underestimated due to the absence of dedicated tracking detector(s).

The detector's branch current, gain stability, and rate-handling capability were investigated by varying gamma intensity at two different GEM operating voltages. At the lower voltage setting, the branch current increased by approximately '12 $\mu$A', accompanied by a 10 $\pm$ 5\% reduction in gain from an initial value of 6.4K. The branch current rose by '15 $\mu$A' at the higher voltage setting, while the gain, initially at 7.2K, decreased by 16.5 $\pm$ 3\%. Notably, these gain reductions were observed only at the highest gamma intensity. This gain reduction is attributed to the increased current flow through the GEM foil segments, leading to a voltage drop across the protection resistance and, consequently, a reduction in the effective GEM voltage. Though direct measurement of detector efficiency has not been carried out in the presence of a gamma background, based on the drop in gain at the mentioned two operating voltages, it is expected that the detector efficiency will remain unaffected.

The average number of digis in the detector shows a linear response with intensity for a gain of around 6.4K, revealing a good gamma rate handling capability up to the highest incident gamma flux of approximately '17.25 MHz/cm$^{2}$'. However, the response of the GEM chamber starts to deviate from linearity beyond this gain for the same gamma-ray intensity. 

Simulation studies using a realistic CBM setup for Au$-$Au collisions at '12 AGeV' and an interaction rate of '10 MHz' predict a maximum digi rate of approximately '120 KHz/pad'. At GIF++, a comparable maximum digi rate of around '125 KHz/pad' has been measured. It might be instructive to have similar measurements with a charged-particle background instead of gammas since their interaction would be closer to the actual CBM environment. This study will be conducted in nucleus-nucleus collisions at the upcoming miniCBM (mCBM) experiment at the Gesellschaft für Schwerionenforschung (GSI) facility in Germany. The combination of the results of this paper and those from mCBM will further enhance our understanding of the rate capabilities of the GEM detector. 

Furthermore, this study does not account for systematic uncertainties in gain values arising from drift gap non-uniformities, local fluctuations, and clustering effects, making it challenging to determine absolute values. Instead, it evaluates the detector's stability in a high gamma-rate environment by examining gain variations as a function of voltage and intensity in a particular detector region.

In summary, the paper illuminates the GEM detector's performance in a high-radiation environment, helps us identify the approximate operational region at extreme incident gamma rates, and attempts to prove its readiness for the upcoming CBM experiment.

\acknowledgments
We gratefully acknowledge the Department of Atomic Energy, Govt. of India, for funding the project. We appreciate the support of the GIF++ team at CERN. We also recognize the help received from the VECC grid computing team for computational work. We also acknowledge the invaluable support of Eraldo Oliveri from CERN, Giuseppe Pezzullo from GIF++, Christian Sturm and Jörg W. Lehnert from GSI, and Pierre-Alain Loizeau from FAIR for their invaluable assistance during the beam test.

%\paragraph{Note added.} This is also a good position for notes to be added
%after the paper has been written.

%\bibliography{journals,phd-references} % Where journals.bib and phd-references.bib are BibTeX databases
\bibliographystyle{unsrt}
% \bibliography{}
\bibliography{manuscript}

\end{document}